\documentclass[iop,apj]{emulateapj}

\usepackage{apjfonts}
\usepackage{psfig}
\usepackage[squaren,Gray]{SIunits}

\newcommand{\AIPS}{{$\cal AIPS\/$}}

\def\aap{A\&A}

\def\apjl{ApJ}
\def\apjs{ApJS}

\def\qir{$q_{\rm IR}$}
\def\lir{$L_{\rm IR}$}

\def\l1.4{$L_{\rm 1.4GHz}$} \def\s1.4{$S_{\rm 1.4GHz}$}

\def\xunits{M$_{\odot}$ (\kelvin\,km\,s$^{-1}$ pc$^2$)$^{-1}$}
\def\kms{km\,s$^{-1}$}
\def\Mgas{$M_{\rm gas}$}
\def\Md{$M_{\rm dust}$}
\def\Td{$T_{\rm dust}$}
\def\Tb{$T_{\rm b}$}
\def\Cplus{[C\,{\sc ii}]}
\def\COtw{$^{12}$C$^{16}$O}
\def\COth{$^{13}$C$^{16}$O}
\def\CeiO{$^{12}$C$^{18}$O}
\def\jonezero{$J\!=\!1\!-\!0$}
\def\jthreetwo{$J\!=\!3\!-\!2$}
\def\jfourthree{$J\!=\!4\!-\!3$}
\def\um{$\mu$m}

\def\gs{\mathrel{\raise0.35ex\hbox{$\scriptstyle >$}\kern-0.6em
\lower0.40ex\hbox{{$\scriptstyle \sim$}}}}
\def\ls{\mathrel{\raise0.35ex\hbox{$\scriptstyle <$}\kern-0.6em
\lower0.40ex\hbox{{$\scriptstyle \sim$}}}}
\def\m@th{\mathsurround=0pt }
\def\eqalign#1{\null\,\vcenter{\openup1\jot \m@th
 \ialign{\strut\hfil$\displaystyle{##}$&$\displaystyle{{}##}$\hfil
 \crcr#1\crcr}}\,}

\begin{document}

\title{{\it HERSCHEL}-ATLAS: A BINARY  HyLIRG PINPOINTING A CLUSTER OF
STARBURSTING PROTO-ELLIPTICALS}

\shorttitle{HyLIRGs amidst distant starbursting cluster}
\shortauthors{Ivison et al.}

\author{R.\,J.~Ivison\altaffilmark{1,2}}
\author{A.\,M.~Swinbank\altaffilmark{3}}
\author{Ian~Smail\altaffilmark{3}}
\author{A.\,I.~Harris\altaffilmark{4}}
\author{R.\,S.~Bussmann\altaffilmark{5}}
\author{A.~Cooray\altaffilmark{6}}
\author{P.~Cox\altaffilmark{7}}
\author{H.~Fu\altaffilmark{6}}
\author{A.~Kov\'acs\altaffilmark{8}}
\author{M.~Krips\altaffilmark{7}}
\author{D.~Narayanan\altaffilmark{9}}
\author{M.~Negrello\altaffilmark{10}}
\author{R.~Neri\altaffilmark{7}}
\author{J.~Pe\~{n}arrubia\altaffilmark{2}}
\author{J.~Richard\altaffilmark{11}}
\author{D.\,A.~Riechers\altaffilmark{12}}
\author{K.~Rowlands\altaffilmark{13}}
\author{J.\,G.~Staguhn\altaffilmark{14}}
\author{T.\,A.~Targett\altaffilmark{2}}
\author{S.~Amber\altaffilmark{15}}
\author{A.\,J.~Baker\altaffilmark{16}}
\author{N.~Bourne\altaffilmark{13}}
\author{F.~Bertoldi\altaffilmark{17}}
\author{M.~Bremer\altaffilmark{18}}
\author{J.\,A.~Calanog\altaffilmark{6}}
\author{D.\,L.~Clements\altaffilmark{19}}
\author{H.~Dannerbauer\altaffilmark{20}}
\author{A.~Dariush\altaffilmark{21}}
\author{G.~De~Zotti\altaffilmark{22,10}}
\author{L.~Dunne\altaffilmark{23}}
\author{S.\,A.~Eales\altaffilmark{24}}
\author{D.~Farrah\altaffilmark{25}}
\author{S.~Fleuren\altaffilmark{26}}
\author{A.~Franceschini\altaffilmark{27}}
\author{J.\,E.~Geach\altaffilmark{28}}
\author{R.\,D.~George\altaffilmark{2}}
\author{J.\,C.~Helly\altaffilmark{3}}
\author{R.~Hopwood\altaffilmark{19,15}}
\author{E.~Ibar\altaffilmark{1,29}}
\author{M.\,J.~Jarvis\altaffilmark{30,31}}
\author{J.-P.~Kneib\altaffilmark{32}}
\author{S.~Maddox\altaffilmark{23}}
\author{A.~Omont\altaffilmark{33}}
\author{D.~Scott\altaffilmark{34}}
\author{S.~Serjeant\altaffilmark{15}}
\author{M.\,W.\,L.~Smith\altaffilmark{24}}
\author{M.\,A.~Thompson\altaffilmark{28}}
\author{E.~Valiante\altaffilmark{24}}
\author{I.~Valtchanov\altaffilmark{35}}
\author{J.~Vieira\altaffilmark{36}}
\author{P.~van~der~Werf\altaffilmark{37}}

\altaffiltext{1}{UK Astronomy Technology Centre, Science and Technology Facilities
  Council, Royal Observatory, Blackford Hill, Edinburgh EH9 3HJ, UK}
\altaffiltext{2}{Institute for Astronomy, University of Edinburgh, Royal Observatory, Blackford Hill, Edinburgh EH9 3HJ, UK}
\altaffiltext{3}{Institute for Computational Cosmology, Durham
  University, South Road, Durham DH1 3LE, UK}
\altaffiltext{4}{Dept of Astronomy, University of Maryland, College Park, MD 20742}
\altaffiltext{5}{Harvard-Smithsonian Center for Astrophysics, 60
  Garden St, Cambridge, MA 02138}
\altaffiltext{6}{Dept of Physics \& Astronomy, University of
  California, Irvine CA 92697}
\altaffiltext{7}{Institut de Radioastronomie Millim\'etrique,
    300 rue de la Piscine, 38406 Saint-Martin d'H\`eres, France}
\altaffiltext{8}{Dept of Astronomy, University of Minnesota, Minneapolis, MN 55414}
\altaffiltext{9}{Steward Observatory,  University of Arizona, Tucson, AZ 85721}
\altaffiltext{10}{INAF, Osservatorio Astronomico di Padova, I-35122 Padova, Italy}
\altaffiltext{11}{Centre de Recherche Astrophysique de Lyon,
    Universit\'e Lyon 1, 9 Avenue Charles Andr\'e, F-69561 Saint Genis
    Laval Cedex, France}
\altaffiltext{12}{Dept of Astronomy, Space Science Building,
    Cornell University, Ithaca, NY 14853-6801}
\altaffiltext{13}{School of Physics \& Astronomy, University of
    Nottingham, University Park, Nottingham NG7 2RD, UK}
\altaffiltext{14}{The Henry A.\ Rowland Dept of Physics \& Astronomy,
    Johns Hopkins University, 3400 N.\ Charles Street, Baltimore, MD 21218}
\altaffiltext{15}{Dept of Physical Sciences, The Open University,
  Milton Keynes, UK}
\altaffiltext{16}{Dept of Physics and Astronomy, Rutgers, The
  State University of New Jersey, Piscataway, NJ 08854-8019}
\altaffiltext{17}{Argelander-Institute for Astronomy, Bonn University,
  Auf dem Huegel 71, 53121 Bonn, Germany}
\altaffiltext{18}{H.\,H.\ Wills Physics Laboratory, University of
  Bristol, Tyndall Avenue, Bristol BS8 1TL, UK}
\altaffiltext{19}{Physics Dept, Imperial College London, South
  Kensington Campus, London SW7 2AZ, UK}
\altaffiltext{20}{Universit\"{a}t Wien, Institut f\"{u}r Astrophysik,
  A-1180 Wien, Austria}
\altaffiltext{21}{Institute of Astronomy, University of Cambridge,
    Madingley Road, Cambridge CB3 0HA, UK}
\altaffiltext{22}{SISSA, via Bonomea 265, I-34136 Trieste, Italy}
\altaffiltext{23}{Dept of Physics and Astronomy, University of
  Canterbury, Private Bag 4800, Christchurch, New Zealand}
\altaffiltext{24}{School of Physics \& Astronomy, Cardiff University,
  Queen's Buildings, The Parade 5, Cardiff CF24 3AA, UK}
\altaffiltext{25}{Dept of Physics, Virginia Polytechnic
    Institute \& State University, MC 0435, 910 Drillfield Drive,
    Blacksburg, VA 24061}
\altaffiltext{26}{School of Physics \& Astronomy, Queen Mary
    University of London, Mile End Road, London E1 4NS, UK}
\altaffiltext{27}{Dipartimento di Astronomia, Universita' di Padova,
  35122 Padova, Italy}
\altaffiltext{28}{Centre for Astrophysics Research, University of Hertfordshire, Hatfield AL10 9AB, UK}
\altaffiltext{29}{Universidad Cat\'olica de Chile, Departamento de
    Astronom\'ia y Astrof\'isica, Vicu\~na Mackenna 4860, Casilla 306,
    Santiago 22, Chile}
\altaffiltext{30}{Astrophysics, Dept of Physics, Keble Road,
    Oxford OX1 3RH, UK}
\altaffiltext{31}{Dept of Physics, University of the Western Cape,
  South Africa}
\altaffiltext{32}{EPFL SB IPEP LASTRO, Observatoire Sauverny, CH-1290
    Versoix, Switzerland}
\altaffiltext{33}{UPMC Univ.\ Paris 06 \& CNRS UMR7095 IAP, F75014 Paris, France}
\altaffiltext{34}{Dept of Physics \& Astronomy, University of British Columbia,
  Vancouver V6T 1Z1, Canada}
\altaffiltext{35}{European Space Astronomy Centre, {\it Herschel} Science
  Centre, ESA, 28691 Villanueva de la Ca\~{n}ada, Spain}
\altaffiltext{36}{California Institute of Technology, 1200 East
  California Boulevard, Pasadena, CA 91125}
\altaffiltext{37}{Leiden Observatory, Leiden University, P.O.\ Box
    9513, NL-2300 RA Leiden, The Netherlands}

\slugcomment{Submitted to The Astrophysical Journal}

\begin{abstract}
  Panchromatic observations of the best candidate HyLIRG from the
  widest {\it Herschel} extragalactic imaging survey have led to the
  discovery of at least four intrinsically luminous $z=2.41$ galaxies
  across a $\approx$100-kpc region -- a cluster of starbursting
  proto-ellipticals. Via sub-arcsecond interferometric imaging we have
  measured accurate gas and star-formation surface densities. The two
  brightest galaxies span $\sim$3\,kpc FWHM in submm/radio continuum
  and CO \jfourthree, and double that in CO \jonezero. The broad CO
  line is due partly to the multitude of constituent galaxies and
  partly to large rotational velocities in two counter-rotating gas
  disks -- a scenario predicted to lead to the most intense
  starbursts, which will therefore come in pairs. The disks have
  $M_{\rm dyn}$ of several $\times 10^{11}$\,M$_{\odot}$, and gas
  fractions of $\sim 40$\%. Velocity dispersions are modest so the
  disks are unstable, potentially on scales commensurate with their
  radii: these galaxies are undergoing extreme bursts of star
  formation, not confined to their nuclei, at close to the Eddington
  limit. Their specific star-formation rates place them $\gs5\times$
  above the main sequence, which supposedly comprises large gas disks
  like these. Their high star-formation efficiencies are difficult to
  reconcile with a simple volumetric star-formation law. $N$-body and
  dark matter simulations suggest this system is the progenitor of a
  B(inary)-type $\approx 10^{14.6}$-M$_\odot$ cluster.
\end{abstract}


\section{Introduction}
\label{intro}

Of the known denizons of the galaxy zoo, hyperluminous infrared (IR)
galaxies (HyLIRGs, $L_{\rm IR}\ge 10^{13}$\,L$_{\odot}$, where the IR
luminosity is measured across $\lambda_{\rm rest}=8$--1000\,$\mu$m)
are amongst the rarest and most extreme. They provide excellent
laboratories with which to confront the most recent hydrodynamic
simulations of isolated and merging galaxies
\citep[e.g.][]{hayward11}. The $L_{\rm IR}$ of a HyLIRG implies a
staggering star-formation rate, $\rm SFR \gs
1000$\,M$_{\odot}$\,yr$^{-1}$ \citep[][for a \citealt{chabrier03}
initial mass function, IMF]{kennicutt98}, unless the IMF is
top-heavy, or there is a substantial contribution to $L_{\rm IR}$ from
a deeply obscured AGN, in which case we are seeing intense star
formation accompanied by the rapid growth of a massive black hole
\citep{alexander08}. Either way, we are witnessing galaxy formation at
its most extreme.

How best can we identify the most luminous star-forming HyLIRGs? A
promising method involves searching amongst the brightest sources
detected in the widest surveys with {\it Herschel}. These comprise
blazars, low-redshift spirals and some intrinsically fainter sources
that have been strongly lensed \citep[e.g.][]{negrello10}. However,
there is also the intriguing possibility that such samples may contain
unlensed and thus intrinsically luminous galaxies -- the rarest
HyLIRGs are predicted to have a space density, $\approx
10^{-7}$\,Mpc$^{-3}$ and may mark the sites of today's clusters
\citep{negrello05, lapi11}.

Of the facilities providing redshifts for bright {\it Herschel}
sources, the 100-m Robert C.\ Byrd Green Bank Telescope (GBT), together
with the ultrawide-bandwidth Zpectrometer cross-correlation
spectrometer and Ka-band receiver, have been amongst the most effective
\citep[e.g.][]{swinbank10, frayer11}. Having added considerably to the
sample of such sources with known redshifts, \citet{harris12}
identified a flat trend of $L'_{\rm CO}$ with \COtw\ \jonezero\
line\footnote{Hereafter, CO refers to \COtw\ unless stated otherwise.}
width. This contrasted with the steep power-law relation between
$L'_{\rm CO}$ and CO line width found by \citet{bothwell13} for
unlensed SMGs, as expected if the most gas-rich galaxies tend to live
in the most massive gravitational potentials.  \citeauthor{harris12}
argued that intrinsically fainter sources require a higher lensing
magnification, $\mu$, to rise above the observational detection
threshold for CO \jonezero, which is approximately constant in
flux. The flat trend seen in the GBT sample is then best interpreted
as lensing magnification acting on a population with intrinsically
steep number counts. Those GBT sources closest to the power-law fit
seen for SMGs -- typically those with the widest lines -- are likely
to be suffering the least lensing magnification. This is where we
might expect to find any HyLIRGs that are lurking amongst the lensed
starbursts.

Our approach here, therefore, is to search for HyLIRGs amongst those
bright {\it Herschel} lens candidates with the broadest CO lines,
starting with the best example in the sample observed with GBT by
\citeauthor{harris12}, HATLAS\,J084933.4+021443 (hereafter
HATLAS\,J084933, with 350-$\mu$m flux density, $S_{\rm 350}= 293$\,mJy),
which displays a line consistent with CO \jonezero\ at $z_{\rm
  LSR}=2.410\pm 0.003$, with a full width at half maximum ({\sc
  fwhm}) of $1180\pm 320$\,\kms\ and $S_{\rm CO1-0}= 0.83\pm
0.19$\,mJy. The amplification predicted by \citeauthor{harris12} for
HATLAS\,J084933 is consistent with unity, $\mu=2\pm 1$.

In the next section we describe an extensive set of observations. We
present, analyse, interpret and discuss our reduced images, spectra
and cubes in \S\ref{results}, finishing with our conclusions in
\S\ref{conclusions}. We adopt a cosmology with $H_0 =
71$\,km\,s$^{-1}$\,Mpc$^{-1}$, $\Omega_{\rm m}=0.27$ and
$\Omega_\Lambda = 0.73$, so 1$''$ equates to 8.25\,kpc at $z=2.41$.

\vspace*{0.4cm}
\section{Observations and data reduction}
\label{observations}

{\it Herschel} imaging, undertaken as part of the wide-field {\it
  H}-ATLAS imaging survey \citep{eales10}, led to the selection of
HATLAS\,J084933 as a potentially distant, lensed starburst: distant,
because it is an example of a so-called `350-$\mu$m peaker', with its
thermal dust peak between the 250- and 500-$\mu$m bands; lensed,
because models of the far-IR/submm source counts suggest that the
majority of objects with $S_{500}>100$\,mJy are expected to be either
local ($z<0.1$) or lensed \citep[e.g.][]{negrello10}, along with the
occasional flat-spectrum radio quasar.

On the basis that it represented the best chance of finding an
intrinsically luminous system rather than a lensed galaxy, following
the arguments laid out in \S\ref{intro}, HATLAS\,J084933 was selected
for observations with the Jansky Very Large Array (JVLA). Here we
detail the JVLA observations and those obtained with other facilities
that followed as a consequence of our initial findings, presenting
these datasets in order of decreasing wavelength.

\subsection{JVLA CO \jonezero\ and 5-GHz continuum imaging}
\label{vlaobs}

Whilst the National Radio Astronomy Observatory's (NRAO's)
JVLA\footnote{This work is based on observations carried out with the
  JVLA. The NRAO is a facility of the NSF operated under cooperative
  agreement by Associated Universities, Inc.} was in its DnC, C, B,
BnA and A configurations, between 2012 January and 2013 January, we
acquired $\approx$30\,hr of Ka-band data, scheduled dynamically to
ensure excellent atmospheric phase and pointing stability. We recorded
two sets of eight contiguous baseband pairs, $1024\times 2$-MHz
dual-polarization channels in total.  We tuned the first set of
basebands to cover \COth\ and \CeiO\ \jonezero. The \COtw\ \jonezero\
transition \citep[$\nu_{\rm rest}=115.271203$\,GHz,][]{mn94} was
placed in the second set of baseband pairs, offseting down by 64\,MHz
to 33.740\,GHz to avoid baseband edges.

Around 4\,hr of A-configuration C-band data were also obtained,
recording $1024\times 2$-MHz dual-polarization channels every 1\,sec
across 4.2--6.5\,GHz, with a small gap. Typically, 0.3\,GHz was lost
to severe radio-frequency interference near 6.1\,GHz, yielding a band
center of 5.1\,GHz (5.9\,cm).

PKS\,J0825+0309 and PKS\,J0839+0319 were observed every few minutes to
determine accurate complex gain solutions and bandpass corrections in
the Ka and C bands, respectively. 3C\,286 was observed to set the
absolute flux density scale, and the pointing accuracy was checked
locally every hour.

The data were reduced using \AIPS. The basebands were knitted together
using the {\sc noifs} and {\sc vbglu} tasks, yielding $uv$ datasets
with two intermediate frequencies, each comprising $512\times 2$-MHz
channels.

For the spectral-line data, these were then imaged in groups of four
channels (71\,km\,s$^{-1}$), with natural weighting. A variety of
Gaussian tapers were used to weight the data, with distances to the 30
per cent point of the Gaussian ranging from 100 to 1,000\,k$\lambda$,
to form a number of cubes with FWHM spatial resolutions ranging from
$0.53''\times 0.51''$ to $2.0''\times 1.8''$. The line-integrated map
shown in Fig.~\ref{fig:hires} has a $1.1''\times 1.0''$ synthesized
beam.

The naturally-weighted 5.1-GHz pseudo-continuum data were imaged,
averaging all of the channels, loosely following the techniques
outlined by \citet{om08}. The resulting map has a spatial resolution
of $0.53'' \times 0.42''$ (PA, 26$^{\circ}$) and an r.m.s.\ noise
level of 2.9\,$\mu$Jy\,beam$^{-1}$ (Fig.~\ref{fig:hires}).
 
\subsection{CO \jthreetwo\ imaging from CARMA}
\label{carmaobs}

We used CARMA\footnote{CARMA construction was derived from the states
  of Maryland, California and Illinois, the James S.\ McDonnell
  Foundation, the Gordon and Betty Moore Foundation, the Kenneth T.\
  and Eileen L.\ Norris Foundation, the University of Chicago, the
  Associates of the California Institute of Technology, and the
  NSF. Ongoing CARMA development and operations are supported by the
  NSF under a cooperative agreement, and by the CARMA partner
  universities.} to observe the CO \jthreetwo\ line ($\nu_{\rm
  rest}=345.795991$\,GHz, redshifted to $\nu_{\rm
  obs}$=101.406\,GHz). Observations were carried out with the 3-mm
receivers and the CARMA spectral-line correlator, with an effective
bandwidth of 3.7\,GHz per sideband and a spectral resolution of
15\,\kms. Four tracks were obtained between 2012 February 13--17 in
the C configuration (15--352-m baselines), with 14 usable antennas and
a total on-source observing time of 13.5\,hr. The nearby quasar,
PKS\,J0757+0956, was used for complex gain calibration. The bandpass
shape and absolute flux calibration (the latter good to $\sim$15\%)
were derived from observations of Mars, PKS\,J0854+2006 and
3C\,84. Using natural weighting, we obtained a FWHM beam size of
$2.1'' \times 1.8''$ (PA, 126$^{\circ}$), with a noise level, $\sigma=
0.3$\,mJy\,beam$^{-1}$ (Fig.~\ref{fig:hires}).

\subsection{CO \jfourthree\ imaging from IRAM PdBI}
\label{pdbiobs}

During 2012 February we obtained data using six 15-m antennas in the
most extended configuration of the Institut de Radioastronomie
Millim\'{e}trique's Plateau de Bure Interferometer (IRAM\footnote{
  IRAM is supported by INSU/CNRS (France), MPG (Germany) and IGN
  (Spain).}  PdBI), with baselines ranging up to 800\,m.

The observing frequency was set to 135.203\,GHz, the redshifted
frequency of the CO \jfourthree\ line ($\nu_{\rm
  rest}=461.04077$\,GHz). The weather conditions were exceptionally
good with a precipitable water vapour of around 2\,mm and $T_{\rm
  sys}\sim 100$\,\kelvin\ on average. PKS\,J0825+0309 and
PKS\,J0909+0121 were used to calibrate the complex gains, while 3C\,84
and MWC\,349 were chosen as bandpass and absolute flux
calibrators. The r.m.s.\ noise level is 0.97\,mJy\,beam$^{-1}$ in
4-MHz-wide spectral channels, for a synthesized beam measuring
$1.0''\times 0.5''$ FWHM (PA, 24$^\circ$). Using only the
line-free channels, the r.m.s.\ noise level is
47\,$\mu$Jy\,beam$^{-1}$.

We also report continuum photometry at 1.36\,mm from observations
obtained throughout 2012 as part of a search for water that will be
reported elsewhere.

\subsection{870-$\mu$m continuum imaging from the SMA}
\label{smaobs}

Approximately 1.4, 4.9 and 1.7\,hr of integration were obtained using
the Submillimeter Array (SMA\footnote{The SMA is a joint project
  between the Smithsonian Astrophysical Observatory and the Academia
  Sinica Institute of Astronomy and Astrophysics and is funded by the
  Smithsonian Institution and the Academia Sinica.}) in its compact,
extended and very-extended array configurations during 2011 December
8, 2012 January 31 and 2012 March 31, respectively (2011B-S044). The
receivers were tuned such that the upper sideband was centered at
345\,GHz (870\,$\mu$m), with 8\,GHz of single-polarization bandwidth
in total. 3C\,84 was used as the bandpass calibrator and Titan was
used for absolute flux calibration. PKS\,J0825+0309 and
PKS\,J0909+0121 were again used to track and check the phase and gain.

A {\sc robust} = 2 weighting scheme resulted in a $0.83''\times
0.67''$ beam FWHM (PA, 86$^{\circ}$) and a noise level,
$\sigma=0.77$\,mJy\,beam$^{-1}$ (Fig.~\ref{fig:hires}).

\subsection{350-$\mu$m continuum imaging with CSO/SHARC-2}
\label{sharcobs}

A total of 80\,min of data were collected at 350\,$\mu$m during
periods of excellent weather between 2012 January 12--13 at the 10.4-m
Caltech Submillimeter Observatory (CSO\footnote{This material is based
  upon work at the CSO, which is operated by the California Institute
  of Technology under cooperative agreement with the National Science
  Foundation (AST-0838261).}), using the SHARC-2 camera
\citep{dowell03}. We took advantage of the open-loop, actuated surface
of the telescope \citep{leong06} and we checked and updated the focus
settings several times each night, resulting in near-Gaussian,
diffraction-limited (8.5$''$ FWHM) beam profiles.  Data were reduced
using {\sc crush} \citep{kovacs08b} and the resulting maps have a
noise level, $\sigma=8$\,mJy\,beam$^{-1}$ (Fig.~\ref{fig:hires}).

\subsection{Continuum imaging from {\it Herschel}}
\label{hershelobs}

The acquisition and reduction of {\it Herschel} \citep{pilbratt10}
parallel-mode data from SPIRE \citep{griffin10} and PACS
\citep{poglitsch10} for the 9-hr Science Demonstration Phase field of
{\it H}-ATLAS \citep{eales10} are described in detail by
\citet{ibar10}, \citet{pascale11} and \citet{rigby11}. Here, we use
the original SPIRE imaging and have obtained much deeper PACS data
from the {\sc ot1} programme, {\sc ot1\_rivison\_1}. We recorded data
simultaneously at 100 and 160\,$\mu$m whilst tracking across the
target at 20$''$\,s$^{-1}$ in ten 3-arcmin strips, each offset
orthogonally by 4$''$. A total of 180\,s on source was acquired for
each of two near-orthogonal scans. These data were tackled with a
variant of the pipeline developed by \citet{ibar10pacs}, reaching
$\sigma\approx 4$ and 7\,mJy at 100 and 160\,$\mu$m, respectively. The
160-$\mu$m PACS image is shown in Fig.~\ref{fig:hires}.

\subsection{Infrared continuum imaging from {\it
    Spitzer}, VISTA and the {\it Hubble Space Telescope}}
\label{NIRimaging}

3.6- and 4.5-$\mu$m images were acquired using the Infrared Array
Camera \citep[IRAC --][]{fazio04} aboard {\it Spitzer}\footnote{This
  work is based in part on observations made with the {\it Spitzer
    Space Telescope}, which is operated by the Jet Propulsion
  Laboratory, California Institute of Technology under a contract with
  NASA.} \citep{werner04} on 2012 June 24 as part of program
80156. The imaging involved a 36-position dither pattern, with a total
exposure time of just over 1\,ks, reaching r.m.s.\ depths of 0.3 and
0.4\,$\mu$Jy at 3.6 and 4.5\,$\mu$m, respectively. Corrected basic
calibrated data, pre-processed by the {\it Spitzer} Science Center,
were spatially aligned and combined into mosaics (shown in
Fig.~\ref{fig:hires}) with a re-sampled pixel size of 0.6$''$ and
angular resolution of 2--2.5$''$, using version 18.5.0 of {\sc mopex}
\citep{makovoz05}.

Images were obtained in $Z, Y , J, H$ and $K_{\rm s}$ as part of
VIKING, a public survey with the 4-m Visible and Infrared Survey
Telescope for Astronomy (VISTA) at Paranal, Chile \citep{emerson10},
with an image quality of 0.9$''$ FWHM, as shown in
Fig.~\ref{fig:hires}.
 
A SNAPshot observation was obtained with the {\it Hubble Space
  Telescope}\footnote{Based on observations made with the NASA/ESA
  {\it Hubble Space Telescope}, obtained at the Space Telescope
  Science Institute, which is operated by the Association of
  Universities for Research in Astronomy, Inc., under NASA contract
  NAS 5-26555. These observations are associated with program 12488.}
{\it (HST)} on 2012 February 08, as part of Cycle-19 proposal 12488,
using Wide-Field Camera 3 (WFC3) with its wide $J$ filter, F110W. The
total exposure time was 252\,s. Data were reduced using the {\sc iraf}
MultiDrizzle package. Individual frames were corrected for distortion,
cleaned by cosmic rays and other artifacts and median combined. The
resulting $\sim2'\times 2'$ image was re-sampled to a finer pixel
scale of 0.064$''$ and is shown in Fig.~\ref{fig:hires}.

\subsection{Optical spectroscopy from Keck}
\label{keckspectrum}

We observed several optical/IR sources in the vicinity of
HATLAS\,J084933 using the DEIMOS spectrograph on the 10-m Keck-{\sc
  ii} telescope\footnote{Data presented herein were obtained at the
  W.\,M.\ Keck Observatory, which is operated as a scientific
  partnership among the California Institute of Technology, the
  University of California, and NASA.  The Observatory was made
  possible by the generous financial support of the W.\,M.\ Keck
  Foundation.} during 2012 March 02. DEIMOS was used with its 600ZD
grating (ruled with 600 lines\,mm$^{-1}$), the GG455 filter and a
1$''$ slit at a central wavelength of 720\,nm. Two positions were
targeted, each with a different PA, for 40 and 75\,min.

\section{Results, analysis and discussion}
\label{results}

\subsection{Basic morphological description}
\label{morph}
 
\begin{figure*}
\centerline{\psfig{file=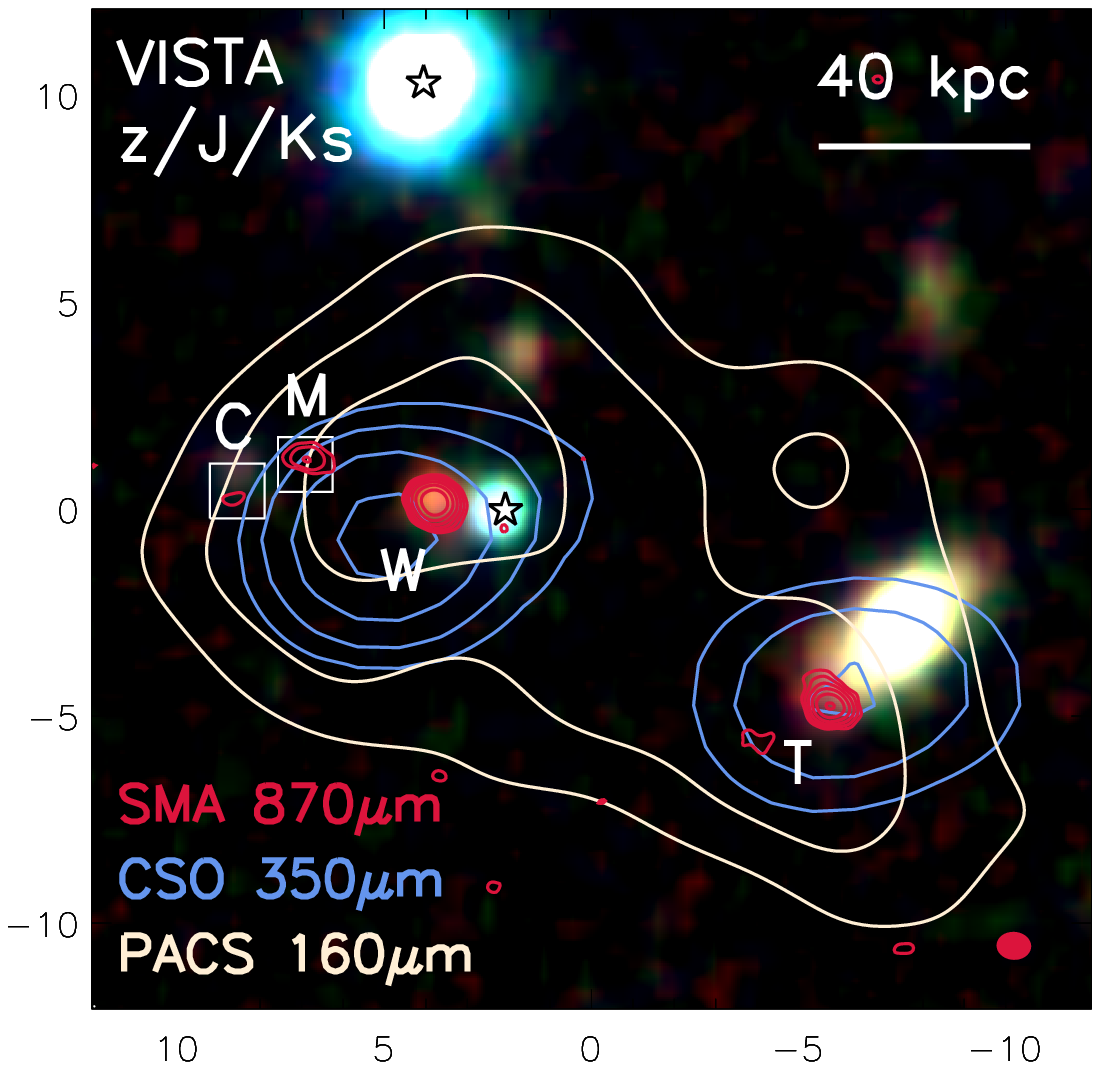,width=3.5in,angle=0}
\hspace{0.1cm}
\psfig{file=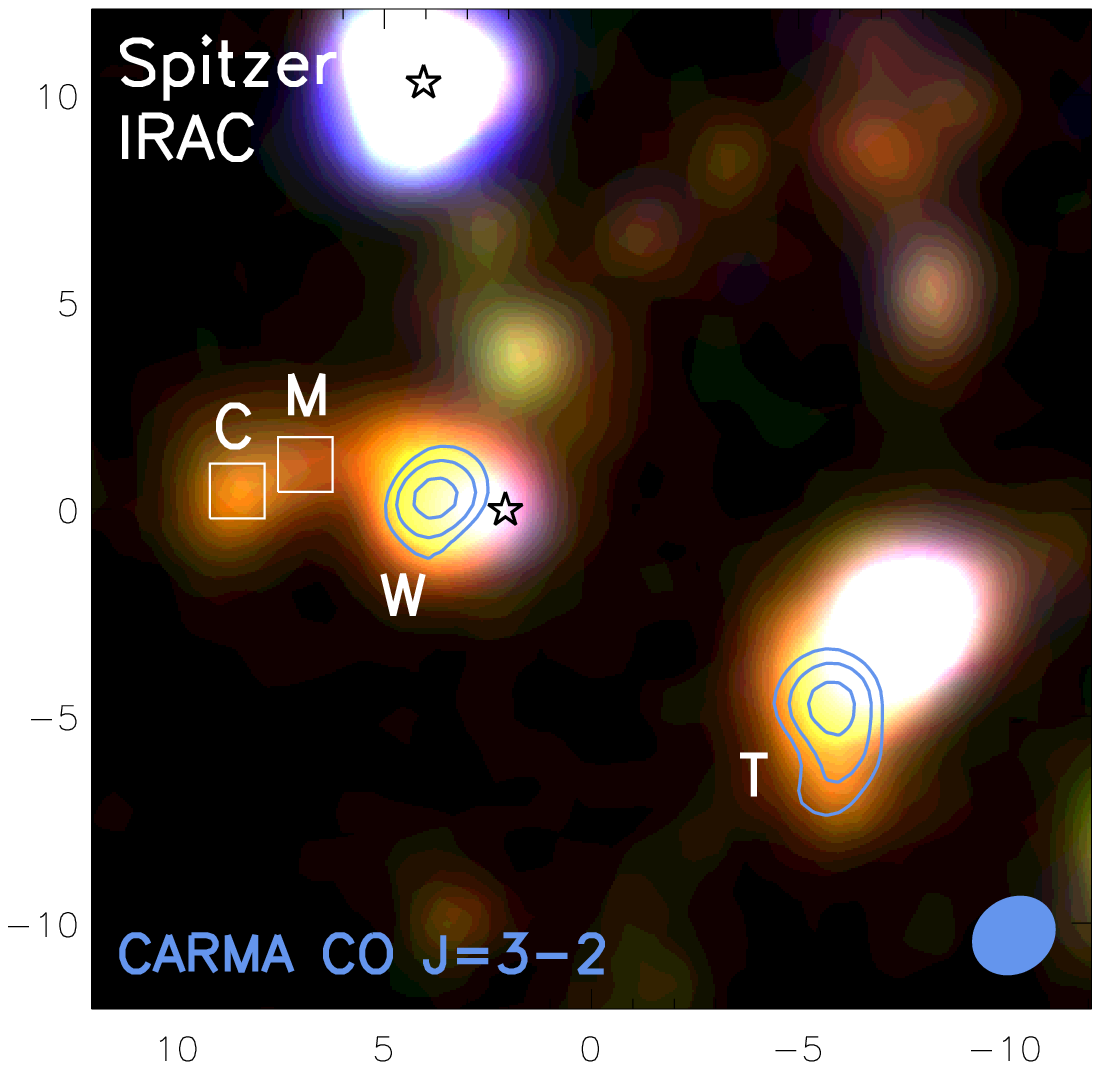,width=3.5in,angle=0}}
\vspace{0.3cm} 
\centerline{\psfig{file=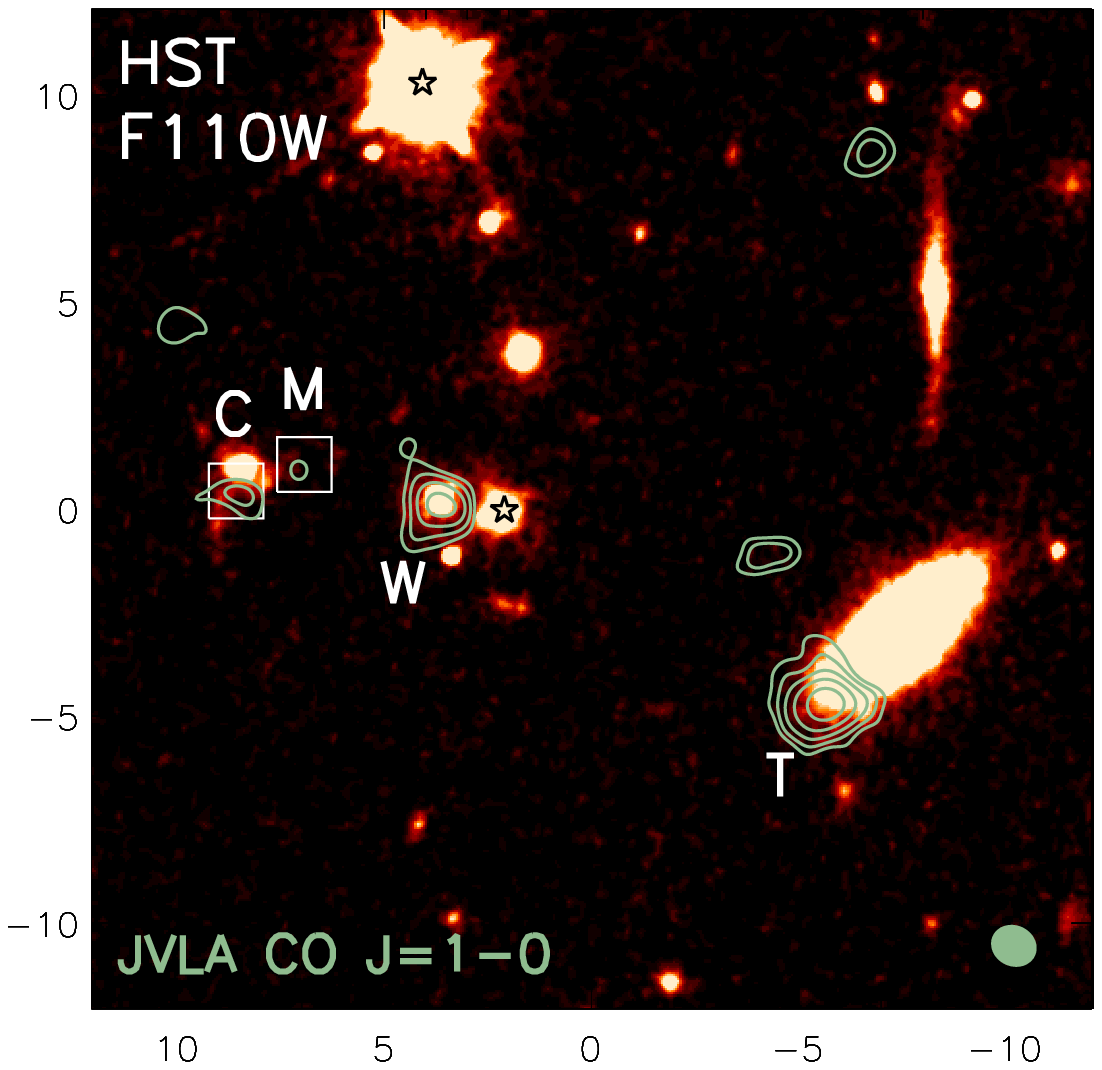,width=3.5in,angle=0}
\hspace{0.1cm}
\psfig{file=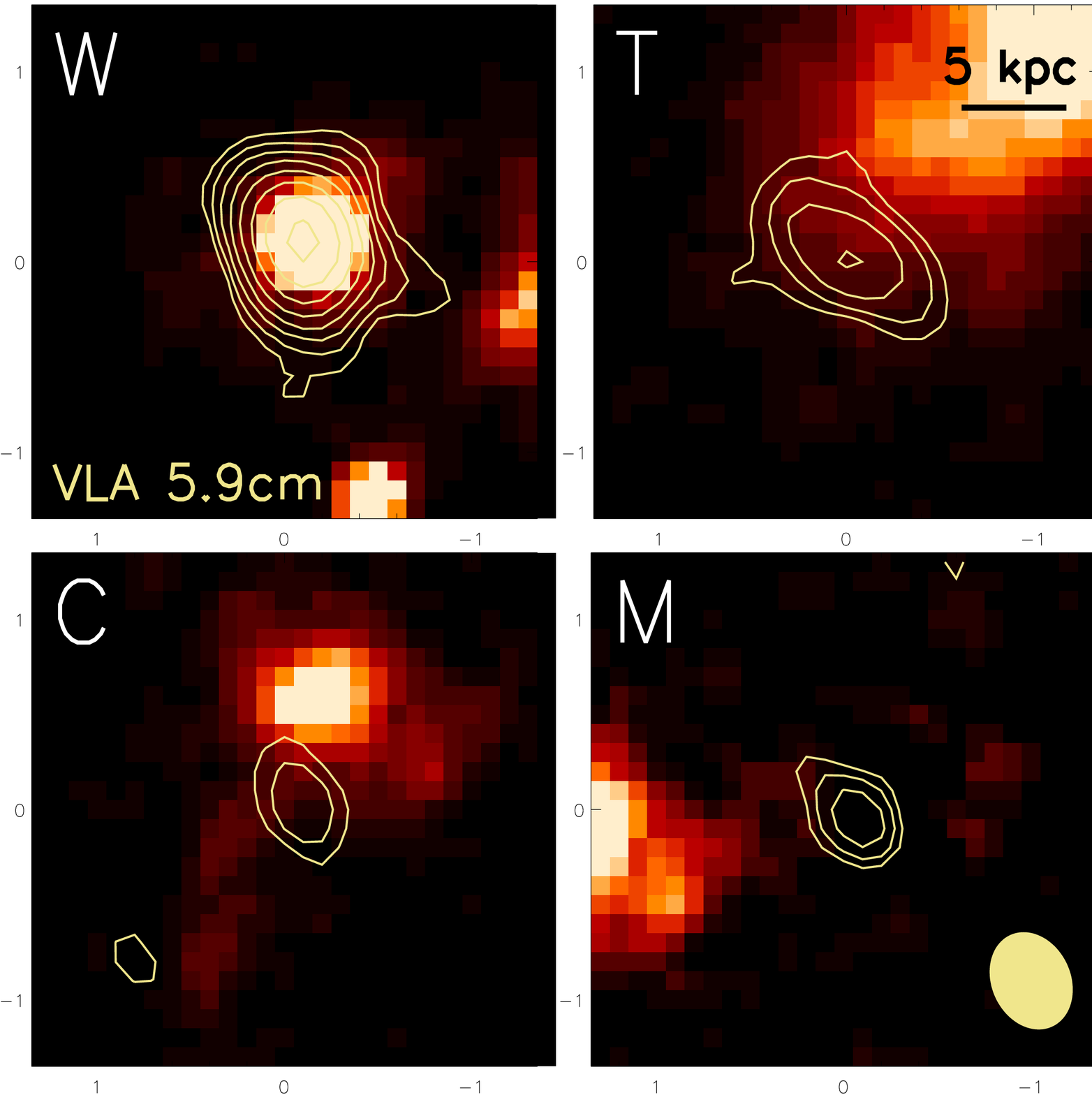,width=3.5in,angle=0}}
\caption{Morphology of HATLAS\,J084933 at moderate-to-high spatial
  resolution. {\it Top left:} 160-, 350- and 870-$\mu$m continuum
  emission from dust, as detected by {\it Herschel}/PACS, CSO/SHARC-2
  and the SMA, superimposed on a three-color representation of the
  VISTA $z, J$ and $K_{\rm s}$ data.  We see two bright components: W
  is centered on a red galaxy, near a blue star; T lies at the end of
  a lenticular galaxy at $z=0.3478$, coincident with its red southern
  tip. Dust emission can also be seen coincident with two further
  galaxies, labelled C and M, to the east of W. Both C and M can also
  seen in continuum at 5.9\,cm (see {\it lower right}), at 1.36\,mm,
  2.21\,mm, and in CO \jonezero\ and \jfourthree. {\it Top right:} CO
  \jthreetwo\ emission superimposed on a three-color representation of
  the {\it Spitzer} IRAC 3.6- and 4.5-$\mu$m imaging and a heavily
  smoothed VISTA $J + H + K_{\rm s}$ image as the blue channel. CO
  emission is coincident with the two submm-bright clumps, W and T. An
  apparent extension south of T is mirrored in the IRAC image,
  suggesting that both are real.  {\it Lower left:} JVLA CO \jonezero\
  imaging superimposed on our {\it HST} WFC3 F110W snapshot, wherein
  we see faint diffraction spikes from the stars.  Distorted,
  low-surface-brightness emission is seen coincident with galaxy C,
  characteristic of an interaction. {\it Lower right:} $\approx
  1.25''\times 1.25''$ stamps centered on M, C, W and T, showing our
  JVLA 5.8-cm continuum imaging as blue contours superimposed on the
  {\it HST} F110W imaging. At the high spatial resolution available
  here via interferometric imaging, we see a consistent morphological
  picture for the emission from cool gas and dust in the
  HATLAS\,J084933 system. The emission is dominated by W and T, which
  are separated from one another by $\sim$85\,kpc in the plane of the
  sky, with significant emission also visible from two fainter
  galaxies, C and M. Individually, W and T are well resolved
  spatially. In all panels, contours are plotted at $-3, 3 \times
  \sigma$, with $\sqrt 2$-spaced increments thereafter, where $\sigma$
  is the local noise level. Beam FWHM are shown as solid ellipses. N
  is up; E to the left; offsets from $\alpha_{2000}=132.3889^\circ,
  \delta_{2000}=2.2457^\circ$ are marked in arcseconds. Known stars
  are labelled.}
\label{fig:hires}
\end{figure*}

At the spatial resolution of SPIRE, HATLAS\,J08493 is essentially
unresolved. However, with only modestly improved spatial resolution,
via deep imaging with PACS at 100--160\,$\mu$m and CSO/SHARC-2 at
350\,$\mu$m, the system is resolved into two sources, referred to
hereafter as W and T (Fig.~\ref{fig:hires}).

Moving to higher spatial resolution, via interferometric imaging with
SMA, IRAM PdBI and CARMA at rest-frame 255--880\,$\mu$m, we see a
consistent morphological picture for the emission from cool dust in
the HATLAS\,J084933 system (Fig.~\ref{fig:hires}). The dust continuum
emission is dominated by W and T, which are separated by 10.5$''$, or
$\sim$85\,kpc in the plane of the sky, with W the brightest of the two
from the optical all the way through to the radio regime. W is amongst
the brightest unlensed SMGs ever observed interferometrically at submm
wavelengths, with T not far behind \citep{gear00, lutz01, younger07,
  younger08, younger09, wang07, aravena10cosmos, smolcic12a,
  smolcic12b}. Amongst the unlensed SMG population, its flux
density has been equalled only by SMM\,J123711+622212 \citep[also
known as GN\,20, $S_{\rm 870}=22.9\pm 2.8$ --][]{pope05, iono06}.

Aided by our interferometric astrometry we can see that at optical/IR
wavelengths the westernmost clump, T, is barely visible as a red
extension to a lenticular galaxy (Fig.~\ref{fig:hires}). W coincides
with the extremely red component of a red/blue pair, separated by
$\approx$2$''$.  Based on experience with SMGs such as
SMM\,J02399$-$0136 (e.g.\ Ferkinhoff et al., in preparation), we
expected to find evidence of an AGN towards the compact blue source,
but found instead that it is a star -- our optical spectroscopy
(\S\ref{keckspectrum}) reveals absorption due to Mg as well as the Na
D doublet.

There is no obvious trend with wavelength for the relative IR/submm
flux density contributions from W and T (listed in
Table~\ref{tab:fluxes}), $S_{\rm W}/S_{\rm T}= 1.22\pm 0.33$, $2.22\pm
0.42$, $1.27\pm 0.19$, $1.32\pm 0.17$, $1.13\pm 0.10$ and $1.23\pm
0.25$ at 100\,$\mu$m, 160\,$\mu$m, 350\,$\mu$m, 870\,$\mu$m, 1.36\,mm
and 2\,mm, respectively. The error-weighted mean for the IR/submm
wavelength regime is $1.22\pm 0.07$.  Moving longward to rest-frame
1.7\,cm, where we expect the emission to be dominated by synchrotron
radiation related to supernova (SN) remnants, HATLAS\,J084933 is also
dominated by component W (Fig.~\ref{fig:hires}), with $S_{\rm
  W}/S_{\rm T}= 2.7\pm 0.7$.

\begin{deluxetable}{rlccl}
\tabletypesize{\scriptsize}
\tablecaption{Continuum flux densities \label{tab:fluxes}}
\tablewidth{0pt}
\tablehead{
\multicolumn{2}{c}{Wavelength} &\multicolumn{2}{c}{$S_{\nu}$} &\colhead{Facility}}
\tablehead{
 \colhead{}                               &\colhead{}&\colhead{W}&\colhead{T}&\colhead{}}
\startdata
   0.88& \um & $8.16\pm 1.37$ & $5.80\pm 0.98$ &$\mu$Jy; VISTA $Z$\\
   1.02& \um & $10.5\pm 1.8$ & $8.65\pm 1.45$ &$\mu$Jy; VISTA $Y$\\
   1.1  & \um  & $13.6\pm 2.3$ & $11.2\pm 1.9$ &$\mu$Jy; F110W\\
   1.25 & \um & $16.8\pm 2.8$ & $13.0\pm 2.2$ &$\mu$Jy; VISTA $J$\\
   1.65& \um & $23.5\pm 4.0$& $19.2\pm 3.2$ &$\mu$Jy; VISTA $H$\\
   2.15& \um & $38.1\pm 6.4$& $26.1\pm 4.4$&$\mu$Jy; VISTA $K_{\rm s}$\\
     3.6 & \um & $67.1\pm 18.6$ & $47.7\pm 17.4$&$\mu$Jy; IRAC\\
     4.5 & \um & $76.6\pm 20.1$ & $45.8\pm 15.2$&$\mu$Jy; IRAC\\
    100 & \um & $23\pm 7$\tablenotemark{a} & $19\pm 7$\tablenotemark{a} &mJy; PACS\\
    160 & \um & $91\pm 12$\tablenotemark{a} & $40\pm 12$\tablenotemark{a} &mJy; PACS\\
    250 & \um & \multicolumn{2}{c}{$242\pm 18$\tablenotemark{b}} &mJy; SPIRE\\
    350 & \um & \multicolumn{2}{c}{$293\pm 22$\tablenotemark{b}} &mJy; SPIRE\\
    500 & \um & \multicolumn{2}{c}{$231\pm 19$\tablenotemark{b}} &mJy; SPIRE\\
    870 & \um & $25\pm 2$ & $19\pm 2$ &mJy; SMA\\
    1.36& mm & $8.3\pm 0.5$&$7.5\pm 0.5$&mJy; PdBI\\ 
     2.21& mm & $1.1\pm 0.1$ &$0.8\pm 0.1$&mJy; PdBI\\
      3.0& mm & $0.33\pm 0.14$ & $3\sigma<0.42$ &mJy; CARMA\\
      8.8& mm & $0.043\pm 0.010$ & $3\sigma < 0.030$ &mJy; JVLA\\
      5.9& cm  & $0.177\pm 0.015$ & $0.065\pm 0.015$ &mJy; JVLA\\
\tableline
\vspace*{-2mm}&&&&\\
             &       &   C   \vspace*{0.5mm}                     & M &\\
\tableline
   0.88& \um & $5.96\pm 1.00$ & -- &$\mu$Jy; VISTA $Z$\\
   1.02& \um & $5.46\pm 0.92$ & --&$\mu$Jy; VISTA $Y$\\
   1.1  & \um  & $7.66\pm 1.29$ & -- &$\mu$Jy; F110W\\
   1.25 & \um & $5.72\pm 0.96$ & -- &$\mu$Jy; VISTA $J$\\
   1.65& \um & $5.09\pm 0.86$& -- &$\mu$Jy; VISTA $H$\\
   2.15& \um & $11.1\pm 1.9$& -- &$\mu$Jy; VISTA $K_{\rm s}$\\
     3.6 & \um & $19.5\pm 5.9$ & --&$\mu$Jy; IRAC\\
     4.5 & \um & $17.7\pm 5.3$ & --&$\mu$Jy; IRAC\\
    870 & \um & $4.6\pm 1.9$ & $6.9\pm 1.6$ &mJy; SMA\\
    1.36& mm & $1.8\pm 0.2$&$1.5\pm 0.2$&mJy; PdBI\\
     2.21& mm & $0.25\pm 0.06$ &$0.31\pm 0.08$&mJy; PdBI\\
      5.9& cm  & $0.022\pm 0.006$ & $0.020\pm 0.005$ &mJy; JVLA
\enddata

\tablenotetext{a}{Errors determined by placing many apertures across
  the map, with 3- and 5-\% calibration uncertainties added in
  quadrature at 100 and 160\,$\mu$m, respectively.}
\tablenotetext{b}{Errors include the contribution due to confusion
  and a 7-\% calibration uncertainty has been added in
  quadrature (Valiante et al., in preparation).}
\end{deluxetable}

Continuum emission from dust is also seen east of W, from a galaxy we
label M in Fig.~\ref{fig:hires}. M is detected securely in 870-$\mu$m,
1.36-mm, 2.21-mm and 5.9-cm continuum (Table~\ref{tab:fluxes}).

Still further to the east, emission can be seen from a very red,
morphologically distorted galaxy -- labelled C in Fig.~\ref{fig:hires}
-- which bears a strong resemblence to the Antennae
\citep[e.g.][]{whitmore95, klaas10}. C is detected weakly in
870-$\mu$m, 1.36-mm, 2.21-mm and 5.9-cm continuum
(Table~\ref{tab:fluxes}).

As we shall see in more detail later (\S\ref{spectral}), C and M are
also detected in our CO \jonezero\ and \jfourthree\ spectral-line
imaging, i.e.\ they also lie at $z=2.41$, alongside W and T.

Both M and C are coincident with IRAC 3.6--4.5-$\mu$m emission,
suggesting that these are not tidal tails or gaseous streams, but
rather significant concentrations of stars, gas and dust -- luminous
SMGs in their own right.

Our deep rest-frame 133--289-nm Keck spectroscopy
(\S\ref{keckspectrum}), with slits covering W, T and C, reveals no
significant line emission from T and C; the only line visible from W,
albeit very faint, is C\,{\sc ii} at 232.6\,nm.

Our CO \jthreetwo\ image from CARMA is shown in Fig.~\ref{fig:hires}.
It is interesting to note that the low-level southern extension to the
\jthreetwo\ emission from T is mirrored at 3.6--4.5-$\mu$m, which
suggests that both are real.

\subsection{Lensing model}
\label{lensmodel}

The lenticular galaxy that lies along the line of sight to component T
lies at $z=0.3478$, as revealed by our optical spectroscopic
observations (\S\ref{keckspectrum}), where the slit was placed along the
major axis of the lenticular. Correcting for the instrumental
resolution, the lines due to Mg and Ca H and K suggest the foreground
galaxy has a velocity dispersion of $190\pm 80$\,\kms. Its stellar
mass is $(1.1\pm 0.4) \times 10^{11}$\,M$_\odot$, adopting the
approach outlined later in \S\ref{stellarmass}. We have modelled this
galaxy with an elliptical ($e=0.3$) mass profile, truncated smoothly
at a radius of 25\,kpc (the exact choice has little effect since the
lensed image is well within this radius). The line of sight to the
$z=2.41$ starburst lies close to the semi-major axis of this
foreground structure and suffers a magnification of $\mu = 1.5\pm
0.2$, where the uncertainty was calculated by varying the model
parameters -- orientation, ellipticity and velocity dispersion -- by
10\%.

Galaxy W suffers no significant lensing magnification, as far as it is
possible to discern from our extensive, wide-field, panchromatic
imaging data (\S\ref{NIRimaging}), which are not easily reconciled
with a massive foreground galaxy group or cluster.

\begin{deluxetable*}{lcccc}
\tabletypesize{\scriptsize}
\tablecaption{Properties of HATLAS\,J084933.4+021443\label{tab:properties}}
\tablewidth{0pt}
\tablehead{
\colhead{Property}&\colhead{W}&\colhead{T}&\colhead{M}&\colhead{C}}
\startdata
R.A.\ (J2000) & 08:49:33.59 &  08:49:32.96 & 08:49:33.80 & 08:49:33.91 \\ 
Dec.\ (J2000) & +02:14:44.6 & +02:14:39.7 & +02:14:45.6 & +02:14:45.0\\
Mean FWHM at 5.9\,cm, 870\,$\mu$m and in CO \jfourthree&$2.9\pm 0.4$\,kpc&$3.8\pm 0.6$\,kpc\tablenotemark{a}&\tablenotemark{b}&\tablenotemark{b}\\
FWHM in CO \jonezero\ /kpc &$7.0\pm 2.1$\,kpc&$6.2\pm 1.4$\,kpc\tablenotemark{a}&\tablenotemark{b}&\tablenotemark{b}\\
log \lir\ /L$_{\odot}$&$13.52\pm 0.04$&$13.16\pm 0.05$\tablenotemark{a}&$12.9\pm 0.2$&$12.8\pm 0.2$\\ 
SFR /M$_{\odot}$\,yr$^{-1}$ \citep[][\citealt{chabrier03} IMF]{kennicutt98}&3400&1500\tablenotemark{a}&800&640\\
\Td\ /{\sc k}&$39.8\pm 1.0$&$36.1\pm 1.1$&\tablenotemark{b}&\tablenotemark{b}\\ 
\qir&$2.30\pm0.08$&$2.53\pm 0.13$&\tablenotemark{b}&\tablenotemark{b}\\  
log \Md\  /M$_{\odot}$&$9.32\pm 0.05$&$9.10\pm 0.05$\tablenotemark{a}&\tablenotemark{b}&\tablenotemark{b}\\
CO \jonezero\ $I_{\rm CO}$ /Jy\,\kms &  $0.49\pm 0.06$ & $0.56\pm 0.07$&$0.057\pm0.013$&$0.079\pm0.014$\\ 
CO \jonezero\ $L^\prime_{\rm CO}$ /$10^9$\,\kelvin\,\kms\,pc$^2$ &  $138\pm 17$ & $157\pm 20$&$16.0\pm3.6$&$22.2\pm3.9$\\  
CO \jonezero\ $L_{\rm CO}$ /$10^6$\,L$_\odot$ &  $6.72\pm 0.82$ &$7.65\pm 0.96$&$0.779\pm0.178$&$0.108\pm0.019$\\  
CO \jonezero\ FWHM /\kms &$825\pm 115$&$610\pm 55$&$320\pm 70$&$250\pm 100$\\
CO \jonezero\ $z_{\rm LSR}$ &$2.4066\pm 0.0006$&$2.4090\pm0.0003$&$2.4176\pm 0.0004$&$2.4138\pm 0.0003$\\
\COth\ and \CeiO\ \jonezero\ $I_{\rm CO}$ /Jy\,\kms &  $3\sigma<0.24$\tablenotemark{c}& $3\sigma<0.24$\tablenotemark{c}&$3\sigma<0.24$\tablenotemark{c}&$3\sigma<0.24$\tablenotemark{c} \\
CO \jthreetwo\ $I_{\rm CO}$ /Jy\,\kms &  $4.08\pm 0.92$ & $5.73\pm 1.13$ &$0.66\pm 0.27$&$1.16\pm 0.37$\\
CO \jthreetwo\ FWHM /\kms &$830\pm 155$&$555\pm 75$&$260\pm 95$&$620\pm 160$\\
CO \jthreetwo\ $z_{\rm LSR}$ &$2.4077\pm 0.0009$&$2.4096\pm 0.0004$&$2.4173\pm 0.0006$&$2.414$\tablenotemark{d}\\
CO \jfourthree\ $I_{\rm CO}$ /Jy\,\kms &  $3.76\pm 0.27$ & $5.10\pm 0.37$&$0.66\pm 0.10$&$0.95\pm 0.15$\\
CO \jfourthree\ FWHM /\kms &$985\pm 115$&$545\pm 30$&$320\pm 50$&$450\pm 90$\\
CO \jfourthree\ $z_{\rm LSR}$ &$2.4068\pm 0.0002$&$2.4090\pm 0.0002$&$2.4178\pm 0.0003$&$2.4149\pm 0.0004$\\
\Tb\ $r_{3-2/1-0}$ & $0.93\pm 0.24$ & $1.14\pm 0.27$&$1.29\pm 0.60$&$1.63\pm 0.59$\\
\Tb\ $r_{4-3/1-0}$ & $0.48\pm 0.07$ & $0.57\pm 0.08$&$0.72\pm 0.20$&$0.75\pm 0.18$\\
$v_{\rm max}$ /\kms&$350\pm 10$&$220\pm 15$&$20\pm 10$&$80\pm 20$\\
$R$ /kpc&$7.8\pm 1.0$&$6.8\pm 0.8$\tablenotemark{a}&\tablenotemark{b}&\tablenotemark{b}\\
log $M_{\rm dyn}$ /M$_{\odot}$&$11.51\pm 0.05$&$11.13\pm 0.05$\tablenotemark{a}&$11.11\pm 0.15$&$11.11\pm 0.20$\\
log $M_{\rm H_2+He}$ /M$_{\odot}$, for $\alpha_{\rm  CO}=0.8$\,\xunits&$11.04\pm 0.05$&$10.92\pm 0.06$\tablenotemark{a}&$10.11\pm 0.09$&$10.25\pm 0.07$\\
SFE /L$_{\odot}$\,M$_{\odot}^{-1}$&370&210 \tablenotemark{a}&760&440\\
log $M_{\rm stars}$ /M$_{\odot}$&$11.38\pm 0.12$&$11.01\pm 0.12$\tablenotemark{a}&$\approx$10&$10.36\pm 0.18$
\enddata

\tablenotetext{a}{~Corrected for $\mu=1.5\pm 0.2$.}
\tablenotetext{b}{~Insufficient S/N to make a useful measurement.}
\tablenotetext{c}{~Adopting $S_{\rm total}/S_{\rm peak}$ as seen for \COtw.}
\tablenotetext{d}{~Fixed.}
\end{deluxetable*} 

\subsection{Spectral energy distributions of W and T}
\label{seds}

\begin{figure}
\centerline{\psfig{file=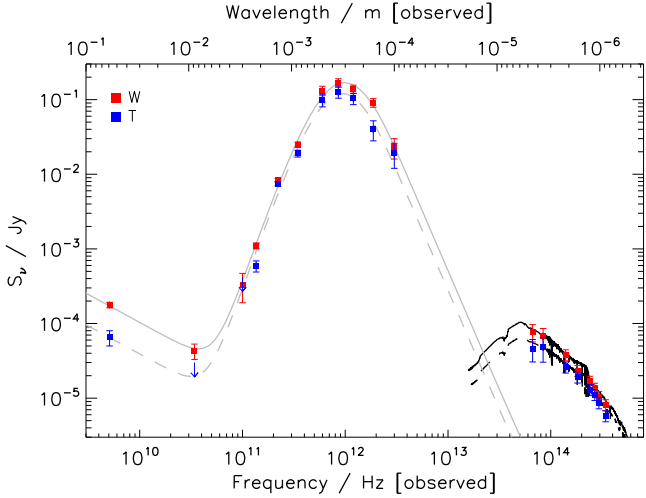,width=3.35in,angle=0}}
\caption{Radio-through-optical SEDs of the dominant components of
  HATLAS\,J084933, W (red points, solid line) and T (blue points, dotted
  line). Our measurements cover the SED peak, constraining \lir\ and
  \Td\ well. The SEDs can be fitted adequately using a combination of
  thermal dust and synchrotron emission components (see \S\ref{seds})
  plus the stellar population modelling described in
  \S\ref{stellarmass}. The key resulting characteristics, \lir, \Td, \Md,
  $M_{\rm stars}$ and \qir\ are listed in
  Table~\ref{tab:properties}.} \label{fig:seds}
\end{figure}

\begin{figure}
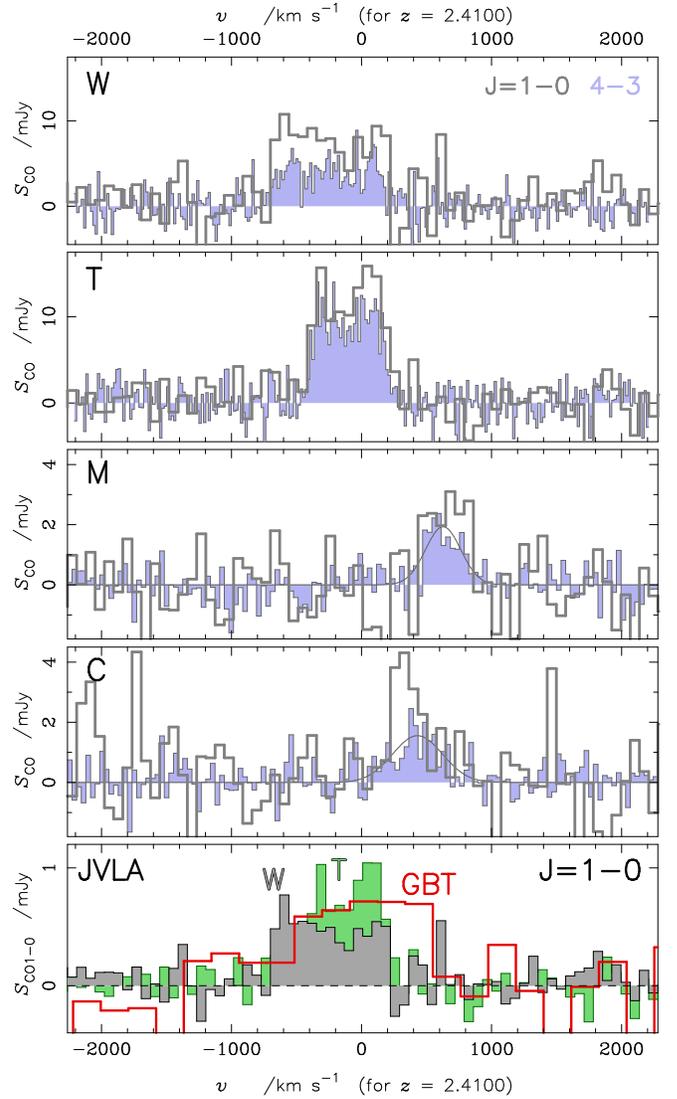

\centerline{\psfig{file=spectrum-w.eps,width=3.4in,angle=270}}
\vspace*{0.0cm}
\centerline{\psfig{file=spectrum-t.eps,width=3.4in,angle=270}}
\vspace*{0.0cm}
\centerline{\psfig{file=spectrum-m.eps,width=3.4in,angle=270}}
\vspace*{0.0cm}
\centerline{\psfig{file=spectrum-c.eps,width=3.4in,angle=270}}
\vspace*{0.0cm}
\centerline{\psfig{file=spectrum-gbt.eps,width=3.4in,angle=270}}
\caption{CO spectra of HATLAS\,J084933. {\it Upper two panels:} CO
  \jonezero\ spectra of galaxies W and T superimposed on solid
  representations of the CO \jfourthree\ spectra (the former scaled by
  16$\times$ to put them on the same \Tb\ scale). The two galaxies
  both have extraordinarily broad CO lines but the profiles are
  noticably different, so they are not lensed images of the same
  galaxy. The profiles of the CO transitions are indistinguishable for
  each individual component; their \Tb\ ratios are also similar
  (Table~\ref{tab:properties}).  {\it Middle two panels:} CO
  \jfourthree\ spectra of galaxies M and C, with line centers
  significantly redwards of W and T. {\it Lower panel:} JVLA CO
  \jonezero\ spectra of W and T alongside the significantly different
  profile from GBT/Zpectrometer \citep{harris12}. Emission at
  velocities redward of W and T -- from M and C and several other
  faint clumps seen in the JVLA data, spanning a $\sim$100-kpc region
  -- can account for the differences between the GBT and JVLA line
  profiles.}
\label{fig:cospectra}
\end{figure}  

Fig.~\ref{fig:seds} shows the SED of HATLAS\,J084933, concentrating on
the brightest components, W and T. Our CSO/SHARC-2 350-$\mu$m imaging
provides evidence that -- at the peak of their SEDs -- the relative
SPIRE contributions of components W and T follow the average ratio
seen in the resolved imaging, $1.22\pm 0.07$, as described in
\S\ref{morph}. Perhaps unsurprisingly, then, we then find they
contribute in roughly the same ratio, $1.55\pm 0.44$, to an immense
overall bolometric luminosity of HATLAS\,J084933, $L_{\rm IR} =
(4.8\pm 1.0) \times 10^{13}$\,L$_{\odot}$
(Table~\ref{tab:properties}), having corrected the contribution from T
for gravitational amplification (\S\ref{lensmodel}).

The IR-through-radio photometry can be described adequately with a
model comprising synchrotron and thermal dust emission, following
\citet{kovacs10}. Table~\ref{tab:properties} summarises a simultaneous
fit to the dominant cold dust temperature, \Td, to \lir\ and \qir\
\citep[as defined by][but where $S_{\rm IR}$ is measured across
$\lambda_{\rm rest}=8$--1000\,$\mu$m, as is \lir]{helou85}, and to
\Md\ (for a characteristic photon cross-section to mass ratio,
$\kappa_{\rm 850}=0.077$\,m$^2$\,kg$^{-1}$ -- \citealt{dunne00,
  dunne01} -- where we fixed the frequency dependence of the dust
emissivity, $\beta$, to be 2.0, which is physically plausible and
consistent with the best-fit value, $2.08\pm 0.15$). We fixed the
synchrotron power-law index, $\alpha=-0.75$, where $S_\nu \propto
\nu^\alpha$, and we also fixed $\gamma=7.2$ -- the power-law index of
a dust temperature distribution appropriate for local starbursts,
$dM_{\rm dust}/dT_{\rm dust} \propto T_{\rm dust}^{-\gamma}$, which is designed
to offer a physically motivated treatment of the Wien side of the
thermal emission spectrum.  Both W and T have $T_{\rm dust}\sim
36$--40\,\kelvin, commensurate with similarly luminous, dusty
starbursts at $z\sim 2$--3, when calculated using a power-law
temperature distribution \citep{magnelli12}.

We are sensitive to AGN activity via our optical imaging and
spectroscopy (\S\ref{keckspectrum}) and our radio/mid-IR imaging
(\S\ref{vlaobs}, \S\ref{NIRimaging}). In relation to the far-IR/radio
flux ratio, W and T have \qir\ = $2.30\pm 0.08$ and $2.53\pm 0.13$,
respectively, spanning (and consistent with) the tight correlation for
star-forming galaxies \citep[\qir\,$\approx 2.4$,
e.g.][]{yun01}. Summarizing, we have no evidence\footnote{Observations
  of higher-$J$ CO lines would be useful in terms of AGN diagnostics.}
that an AGN contributes significantly to the \lir\ of HATLAS\,J084933,
though of course we have no evidence of absence since even a powerful
AGN can be hidden very effectively by the quantities of dust-rich gas
in this system.

If contributions to $L_{\rm IR}$ from AGN are small, then W and T are
generating stars at a rate of $\approx$3400 and
$\approx$1500\,M$_{\odot}$\,yr$^{-1}$, respectively, for a
\citet{chabrier03} IMF \citep[][though see
\citealt{dwek11}]{kennicutt98}. The estimates of \lir\ for C and M
given in Table~\ref{tab:properties} are scaled from their submm
photometry relative to W and T, assuming the same SED shape. This
suggests that both are close to being HyLIRGs in their own right,
$\approx 7\times 10^{12}$\,L$_\odot$.

Before closing our SED discussion, we add a word of caution regarding
our estimated IR luminosities. If W and T lie in the core of a
proto-cluster, as we shall see in what follows, then lower-luminosity
star-forming cluster galaxies may contribute to the flux densities
measured by the relatively large SPIRE beam at 250--500\,$\mu$m
\citep[see][]{negrello05}. Although the calibration cannot compare
with that of {\it Herschel}, SHARC-2 recovers less flux than SPIRE at
350\,$\mu$m, a possible manifestation of the proto-cluster.

\subsection{Stellar masses}
\label{stellarmass}

The rest-frame ultraviolet-through-near-IR SEDs of W and T are
illustrated in Fig.~\ref{fig:seds}, alongside the far-IR--submm--radio
data. Flux densities are listed in in Table~\ref{tab:fluxes}. Our
high-quality {\it HST}, VIKING and IRAC images at 0.88--4.5\,$\mu$m
are compromised to a certain extent by the unfortunate super-position
of a star and a lenticular galaxy near W and T, respectively, but the
considerable uncertainties associated with estimating stellar mass
dominate over the photometric uncertainties \citep[see,
e.g.,][]{hainline11, michalowski12}.

For components W, T and C we find best-fitting monochromatic
rest-frame $H$-band luminosities, $\log L_{\rm H}$, of 11.82, 11.46
and 11.12\,L$_{\odot}$, respectively. Stellar masses were determined
using {\sc magphys} with the most recent version of the stellar
population models of \citet*{bc03}, a \citet{chabrier03} IMF and the
SED prior distributions described by \citet*{dacunha08}, following
\citet{rowlands12}. The models encompass a wide range of exponentially
declining star-formation histories (SFHs), with bursts superimposed
\citep[see][for details]{dacunha08}. The implied stellar masses are
$\log M_{\rm stars} = 11.38^{+0.11}_{-0.13}, 11.01^{+0.12}_{-0.11}$
and $10.36^{+0.18}_{-0.17}$\,M$_\odot$ for W, T and C, where the
errors are derived from the marginalised probability density function,
which incorporates the uncertainties in the SFH\footnote{Stellar mass
  estimates are generally robust to changes in SFH
  \citep[e.g.][]{ilbert10, ilbert13, pforr12}.} and
photometry. Systematic shifts in stellar mass of $\approx0.2$\,dex
result from adopting a Salpeter IMF or ignoring thermally pulsating
stars on the asymptotic giant branch, and the magnitude of photometric
contamination by buried AGN could well be similar, so we estimate that
our stellar mass estimates are accurate to $\approx0.3$\,dex.

Emission from M can be seen at 3.6--4.5\,$\mu$m but the photometric
uncertainties due to blending with C and W are large so we state only
that its stellar mass must be considerable, $\approx
10^{10}$\,M$_{\odot}$.

There is little sign of the power-law SEDs that can betray the
presence of AGN, supporting the lack of AGN spectral features reported
in \S\ref{morph}, though of course this does not rule out the presence
of a deeply buried, highly obscured AGN. 

\subsection{Spectral-line characteristics}
\label{spectral}
  
Fig.~\ref{fig:cospectra} shows the JVLA CO \jonezero\ spectra of W and
T alongside our IRAM CO \jfourthree\ spectra.  These spectral-line
data bring us velocity information for the first time, enabling us to
see that although W and T do lie at approximately the same redshift
and share similar line profiles and \Tb\ ratios, they have
significantly different line widths (Table~\ref{tab:properties}).
Each one is thus a distinct, extraordinarily luminous starburst. While
the CO line profile of T is typical of SMGs
\citep[e.g.][]{bothwell13}, that of W is broader than most, with
$\approx$1000\,\kms\ {\sc fwzi}. Both profiles are flat-topped or
perhaps double-peaked; they can not be described well by a single
Gaussian.

The CO \jonezero\ fluxes of W and T measured in our JVLA data sum to
$I_{\rm CO1-0}=1.05\pm 0.09$\,Jy\,\kms. At first sight this is
consistent with the \citeauthor{harris12} GBT measurement, $1.04\pm
0.37$\,Jy\,\kms. However, the CO \jonezero\ line profiles observed by
JVLA and GBT appear noticeably different in Fig.~\ref{fig:cospectra}:
emission in the GBT spectrum continues significantly redward (by
$\approx$500\,\kms) of the emission that can attributed to W and T.

Looking at our IRAM PdBI imaging -- see Fig.~\ref{fig:iram} -- after
subtracting the significant continuum emission (several submm
continuum flux densities are listed in Table~\ref{tab:fluxes}), we see
that both C and M are detected securely in CO \jfourthree\ (at 9.0 and
7.2\,$\sigma$, respectively; we also see 3.6--4.5-$\mu$m, submm and
radio continuum emission -- recall Fig.~\ref{fig:hires}).  These
galaxies are therefore unquestionably real and they share
approximately the same redshift as components W and T. Their positions
are listed in Table~\ref{tab:properties}, along with those of the two
brighter components. Their CO emission is centered $\approx$500\,\kms\
redward of W and T.

As well as C and M, a number of other faint emission features can seen
in the integrated CO \jonezero\ image (Fig.~\ref{fig:hires}). Summing
C, M and these faint clumps leads to tolerable agreement between the
GBT and JVLA line profiles. Although we must add $0.87\pm
0.19$\,Jy\,\kms\ to $I_{\rm CO}^{\rm JVLA}$, we must also increase
$I_{\rm CO}^{\rm GBT}$ by $\approx$15\% to account for the
$\sim$8-per-cent lower flux density assumed for 3C\,286 by
\citet{harris12}, and for losses due to attenuation of the faint
clumps by the GBT primary beam. The agreement\footnote{Note that the
  low-level GBT spectrometer baseline (at negative velocities relative
  to HATLAS\,J084933) may have caused an underestimate of $I_{\rm
    CO}^{\rm GBT}$.}  between the total GBT and JVLA line intensities
remains consistent to within $\approx 1\sigma$ and the profile
measured by the FWHM $\approx 22''$ GBT primary beam is thereby
reconciled with the JVLA data.

\subsection{Resolved imaging of the gas and dust}
\label{sizes}

Our interferometric images from JVLA, IRAM PdBI and the SMA have a
spatial resolution of $\sim 0.5''$ FWHM, or $\sim 4$\,kpc. They allow
us to compare the sizes of the regions responsible for emission from
the total molecular gas reservoir (via CO \jonezero\ emission), from
the star-forming gas (via CO \jfourthree), from cool dust produced by
SNe in regions of recent star formation (via submm continuum emission)
and from relativistic electrons spiralling in or near recent SN
remnants (via radio continuum emission).

We find that emission from W and T is resolved in all these wavebands,
with total flux densities significantly higher than the peak flux
densities. In Table~\ref{tab:properties} we quote their deconvolved
{\sc fwhm}\footnote{We use half-light radii to calculate the area
  appropriate for gas and star-formation surface densities, alongside
  a $0.5\times$ correction to the SFR or gas mass. We use FWHM simply
  as a convenient way to compare sizes in several wavebands.}  sizes
at 5.9\,cm, 870\,$\mu$m and in CO \jfourthree\ and \jonezero, as
determined with 2-D Gaussian fits.

In the absence of significant AGN-powered activity, radio continuum
observations trace regions where massive stars have recently been
formed, with no obscuration. Of the tracers available to us, we might
anticipate the radio emission to be the most compact, as indeed is
measured. At 870\,$\mu$m we see no evidence of a more extended
emission component, heated by older stars -- the deconvolved sizes of
W and T are consistent, as measured in 5.9-cm and 870-$\mu$m
continuum, as one might expect given the far-IR/radio correlation
(Table~\ref{tab:properties}). The CO \jfourthree\ FWHM are slightly
larger, as seen for local ULIRGs \citep[e.g.][]{wilson08}, but are
consistent with the radio and submm values to within the
uncertainties.
   
The error-weighted average of the 5.9-cm and 870-$\mu$m continuum and
CO \jfourthree\ FWHM measurements is $\rm (3.4\pm 0.4) kpc \times
(2.4\pm 0.4)\,kpc$ (PA, $9\pm 15^{\circ}$) for W; for T, before
correcting for the amplification, we measure $\rm (5.6\pm
0.7)\,kpc\times (3.5\pm 0.7)\,kpc$ (PA, $48\pm 16^{\circ}$).  In CO
\jonezero, however, W and T have considerably larger deconvolved sizes
than those measured in the submm or radio continuum, as seen for SMGs
generally \citep[e.g.][]{ivison11, riechers11b}.

Summarising, we have powerful, direct evidence of starbursts covering
$\approx 3$--4\,kpc FWHM, with still larger reservoirs of gas
available to fuel future star formation, on scales of $\approx
6$--7\,kpc FWHM.

The large sizes measured in CO have been mentioned alongside the
relatively high $L_{\rm [C II]}$/\lir\ ratios measured towards a
number of SMGs \citep[e.g.][]{stacey10}. Both have been taken as
evidence that SMGs form stars across larger spatial scales than the
compact, nuclear events seen in local ULIRGs. 

\subsection{Dynamical characteristics}
\label{dynamics}

\begin{figure*}
  \centerline{\psfig{file=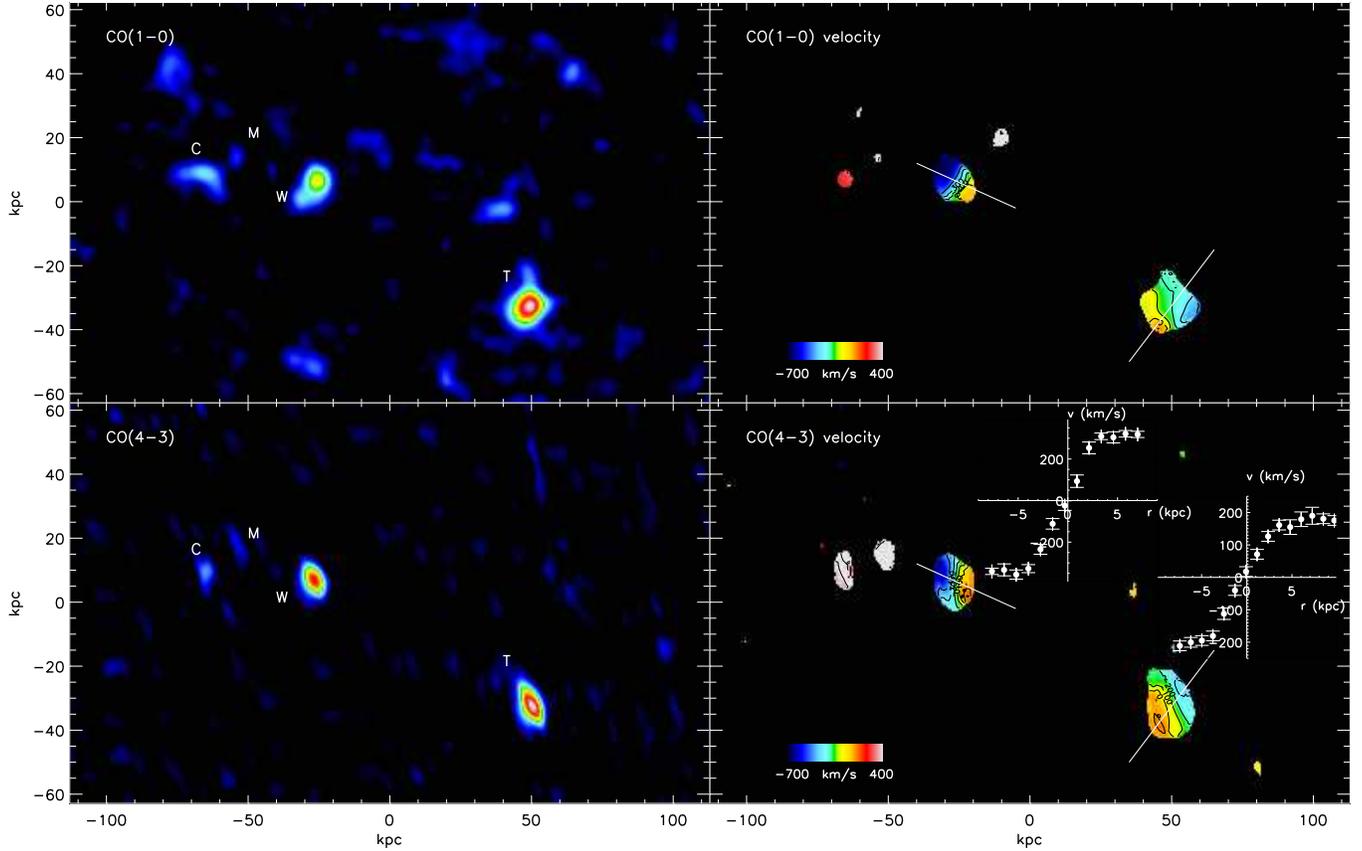,width=7.2in,angle=0}}
  \caption{CO \jonezero\ and \jfourthree\ imaging of
    HATLAS\,J084933. {\it Left:} Velocity-integrated
    continuum-subtracted CO emission from our JVLA and PdBI
    observations.  In both panels, we mark the positions of the
    brightest four galaxies, C, M, W and T. {\it Right:} Dynamical
    maps of the galaxies in CO \jonezero\ and \jfourthree. To
    construct these velocity fields, we fit the CO emission in each
    cube at each pixel, recording the intensity, velocity and line
    width where $\rm S/N>5$.  North is up; East is left. {\it Inset:}
    One-dimensional rotation curves for W and T, extracted along the
    major kinematic axis identified in the modeling.  Both W and T
    have rotation curves which resemble disks; indeed, W and T are
    best described as counter-rotating disks, which \citet*{mihos96}
    predicted would lead to the most intense starbursts.}
\label{fig:iram}
\end{figure*}   

\begin{figure*}
\centerline{\psfig{file=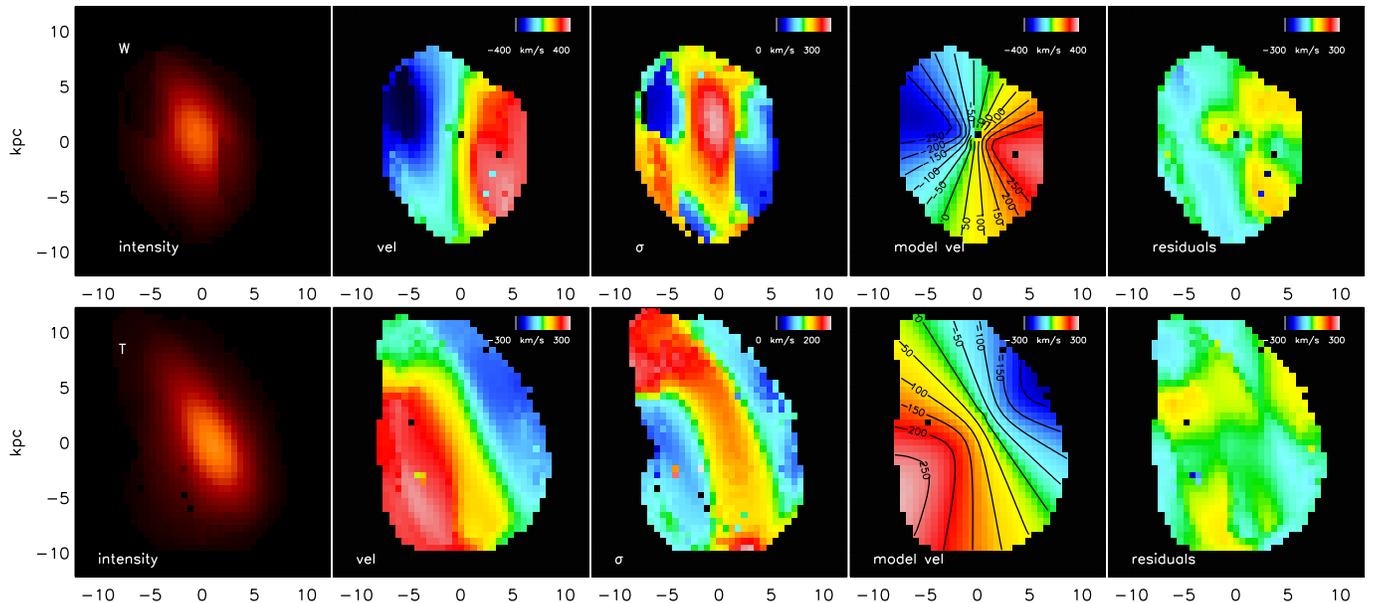,width=7.1in,angle=0}}
\caption{CO \jfourthree\ imaging of galaxies W and T, showing {\it
    left to right} their intensities, resolved velocity profiles,
  line-of-sight velocity dispersions ($\sigma$), our best-fit
  kinematic model (see text) and the residuals.  The color scale for
  the velocity fields is shown in each panel.  We note that there is
  evidence for lensing shear in the velocity field and line-of-sight
  velocity dispersion for T.}
\label{fig:model}
\end{figure*}   
   
The most remarkable characteristics of HATLAS\,J084933 are revealed by
the superlative spatial resolution (\S\ref{sizes}), velocity
resolution and sensitivity of our JVLA CO \jonezero\ and IRAM PdBI CO
\jfourthree\ data, as shown in Fig.~\ref{fig:iram}.

To measure the 2-D velocity structure, we fit the CO emission lines,
pixel by pixel, in each of the cubes. Initially, we attempt to
identify an emission line in $0.2''\times 0.2''$ regions, increasing
this slowly until we reach $\rm S/N>5$, at which point we fit the CO
profile, allowing the intensity, centroid and width to vary.
Uncertainties are calculated by perturbing each parameter
independently, allowing the remaining parameters to find their optimum
values, until $\Delta\chi^2=1$ is reached.

We model the two-dimensional velocity fields of W and T by
constructing and fitting two-dimensional disk models using an arctan
function, $v(r)=2\,\pi^{-1}\,v_{\rm asym}\,{\rm arctan}\,(r/r_t)$, to
describe the kinematics \citep[where $v_{\rm asym}$ is the asymptotic
inclination-corrected rotational velocity and $r_{\rm t}$ is the
effective radius at which the rotation curve turns over --
see][]{courteau97}. We fit six free parameters to the disk velocity
field: $v_{\rm asym}$, $r_{\rm t}$, $x,y$ centroid, PA and inclination
to the sky. The $x,y$ centroid and $r_{\rm t}$ must be within the
field of view and the maximum circular velocity must be $<2\times$ the
maximum velocity seen in the data. Quoted uncertainties reflect the
range of acceptable models from all of the attempted model fits.

Our velocity cubes reveal that the wide CO line that led to this
system being selected as a candidate HyLIRG is due partly to the
multitude of constituent galaxies and partly to large regular,
systematic rotational velocities in the two brightest constituent
galaxies. Both W and T are resolved into counter-rotating gas disks --
the scenario that many studies have predicted should lead to the most
intense starbursts \citep[e.g.][]{mihos94, mihos96, taniguchi98,
  borne00, bekki01, dimatteo07, salome12}.

As shown in Fig.~\ref{fig:model}, we find acceptable fits for W and T
with disk inclinations of $56\pm 10^\circ$ and $49\pm 10^\circ$, at
PAs of 26 and $145^\circ$, with $v_{\rm asym}=415$ and 370\,\kms,
respectively. Component T is sheared slightly by the foreground lens,
as evident from the velocity and $\sigma$ maps. Both galaxies show
small-scale (30--60-\kms, r.m.s.) deviations from the best-fit model.
The dynamical masses\footnote{$R$ was taken to be twice the
  deconvolved half-light radius in CO \jonezero.} of W and T are
listed in Table~\ref{tab:properties}: several $\times
10^{11}$\,M$_{\odot}$, where $M_{\rm dyn}=R\,v_{\rm max}^2/{\rm sin}^2
(i)$. Their velocity dispersions, $\sigma$, corrected for the
instrumental velocity resolution, are $\approx 65$\,\kms.

M has no strong velocity gradient. Its $v_{\rm max}$ is 20\,\kms\
(40\,\kms, peak to peak), $\sigma = 140\pm 15$\,\kms\ for both CO
\jonezero\ and \jfourthree, and $M_{\rm dyn}\approx 1.3\times
10^{11}$\,M$_{\odot}$. For C, in CO \jonezero\ we find $v_{\rm max}=
80$\,\kms\ (160\,\kms, peak to peak), $\sigma = 130\pm 20$\,\kms\ and
$M_{\rm dyn}\approx 1.3\times 10^{11}$\,M$_{\odot}$ (in CO
\jfourthree\ we find slightly higher values for $\sigma$ and $M_{\rm
  dyn}$).

\subsection{Star formation in disks off the main sequence}
\label{ms}

For main-sequence galaxies, star formation is not confined to a
compact starbursting nucleus \citep[e.g.][]{elbaz11}; at $z\sim 2.4$,
the specific SFR is approximately 2.8\,Gyr$^{-1}$. W and T have sSFRs
of 14.2 and 14.7\,Gyr$^{-1}$, so their starburstiness,\footnote{We
  cannot determine the $IR8$ starburst indicator \citep{elbaz11} for
  HATLAS\,J084933 since we lack sufficiently deep rest-frame 8-$\mu$m
  imaging.} $\rm sSFR/sSFR_{MS}\gs 5$, placing them well above the
main sequence. As we saw in \S\ref{sizes} and \S\ref{dynamics} --
their star formation is not centrally concentrated as would be the
case in local starbursts. Instead, W and T are undergoing widespread
star formation in disks that are supported partly by rotation, as
envisaged for main-sequence galaxies, and share many of the properties
of massive star-forming disks \citep[e.g.][]{fs09}. Some other factor
-- presumably their interaction -- has taken them into the realm of
sSFR reserved for starbursts.

For the $Q$ parameter\footnote{Usually ascribed to a seminal paper on
  the stability of stellar disks, \citet{toomre64}, which contains no
  mention of $Q$.} \citep{toomre64, gl65}, which describes the
stability of disks and which has been found to tend towards unity in
stable situations \citep[in models with and without feedback
--][]{hopkins12}, we find values of $\approx 0.35$ for both W and T,
with a plausible range that falls just short of unity. This suggests
the disks are unstable and that the gas therein is prone to condense
and form stars on a timescale shorter than the rotational period.

\subsection{Gas properties and $\alpha_{\rm CO}$}
\label{gasproperties}

The brightness temperature (\Tb) ratios measured for both W and T are
higher than expected (Table~\ref{tab:properties}), where we usually
see $r_{3-2/1-0}\sim 0.52\pm 0.09$ and $r_{4-3/1-0}\sim 0.41\pm 0.07$
for SMGs \citep[][]{harris10,ivison11,bothwell13}, which suggests that
the thermalised CO \jthreetwo\ filling factors may be close to unity
and that we may have found similar sizes in CO \jonezero\ and
\jthreetwo\ if our spatial resolution in the latter had been 2$\times$
higher.

For a CO to H$_2$+He conversion factor, $\alpha_{\rm
  CO}=0.8$\,\xunits, as commonly adopted for IR-luminous starbursts
where the gas is not in virialized individual clouds
\citep{bolatto13}, the gas masses we determine for W, T, C and M are
110, 83.7, 12.8 and $17.8 \times 10^9$\,M$_\odot$ (corrected for
lensing).

If we were instead to adopt the all-sample average gas/dust mass ratio
for nearby galaxies \citep[\Mgas/\Md\ = $91^{+60}_{-36}$
--][]{sandstrom13}, we would determine gas masses for W and T of 19.0
and $11.5 \times 10^{9}$\,M$_\odot$, yielding gas/dust-motivated
$\alpha_{\rm CO}$ values of 1.4 and 1.1\,\xunits, respectively.  Due
to the uncertainty in $\kappa_{850}$, etc., we regard these as
accurate to a factor $2\times$ at best. For an environment with
sub-Solar metallicity, for example, $\alpha_{\rm CO}$ would rise in
line with \Mgas/\Md.

A dynamically-motivated estimate of $\alpha_{\rm CO}$ can be made by
determining the difference between the dynamical and stellar masses in
each system, since this represents the total plausible mass of
gas\footnote{Bearing in mind that any AGN contribution to the
  rest-frame near-IR luminosity will have led to an over-estimate of
  stellar mass.}. Via this method, we find $\alpha_{\rm CO}$ = 0.6 and
0.3\,\xunits\ for W and T, respectively, where the uncertainty,
$\pm0.3$\,dex, is due almost entirely to the notoriously difficult
estimate of stellar mass (\S\ref{stellarmass}).

Finally, we can estimate $\alpha_{\rm CO}$ using its relationship with
gas-phase metallicity and CO surface brightness, as developed by
\citet{narayanan12b}, exploiting the resolved imaging in CO \jonezero\
acquired here. For Solar metallicity \citep{chapman05} and the sizes
that have been measured accurately in our JVLA imaging (\S\ref{sizes};
Table~\ref{tab:properties}) we find $\alpha_{\rm CO}$ = 0.8 and
0.7\,\xunits\ for W and T, respectively.

Our mean is $\alpha_{\rm CO}=0.8\pm 0.4$\,\xunits, in line with the
value usually adopted for starburst galaxies, with a range similar to
that reported by \citet{ds98}, 0.3--1.3\,\xunits. To reconcile our
dynamical measurement of $\alpha_{\rm CO}$ with 3.2\,\xunits, which
\citet{genzel10} argue is appropriate for all disks, regardless of
redshift, would require a buried AGN to dominate the near-IR light,
causing us to over-estimate $M_{\rm stars}$.

Adopting our best estimate for $\alpha_{\rm CO}$, we find that W and T
are both approaching the accepted criterion for maximal starbursts,
with star-formation efficiencies, \lir/\Mgas, of 370 and
210\,L$_\odot$\,M$_\odot^{-1}$. C and M are higher still, though with
considerable uncertainties. The Eddington limit is believed to be
around 500\,L$_\odot$\,M$_\odot^{-1}$; above this level, the radiation
pressure should quickly expel the gas via radiation pressure from
young, massive stars on dust \citep{scoville04}.
 
Feedback due to the aforementioned radiation pressure will lengthen
the total duration of star-formation episodes, or lead to a series of
short bursts. Ignoring this, the gas-consumption timescale implied for
the galaxies comprising HATLAS\,J084933 is $\tau_{\rm gas}\sim$
\Mgas/SFR $\approx$ 15--45\,Myr, consistent with an interaction-driven
period of enhanced star formation \citep*{scoville86}, as already
hinted by the morphologically tortured state of C.

The stellar masses determined in \S\ref{stellarmass} for W, T and C
suggest gas fractions, $f_{\rm gas}=M_{\rm gas}/(M_{\rm gas}+M_{\rm
  stars}= 36$--45\%. These gas fractions are the top end of the range
that \citet*{narayanan12} showed to be difficult to reconcile with
cosmological simulations.

\subsection{Relevance for star-formation laws}
\label{sk}

Data of the quality gathered here have the potential to test whether
the rules and prescriptions developed to describe star formation in
the local Universe \citep[e.g.][]{schmidt59, kennicutt89} can be
applied to to the interstellar medium of distant, gas-rich starbursts
like HATLAS\,J084933. Are the data consistent with a simple,
volumetric star-formation law in which the SFR is $\sim$1\% of the
molecular gas mass per local free-fall time, as argued by
\citet*{krumholz12}? Are their SFRs more closely related to the
orbital periods of their entire galactic disks
\citep[e.g.][]{genzel10, daddi10}, or to the quantity of gas above
some density threshold \citep[e.g.][]{lada10, heiderman10}?

\begin{figure}
\centerline{\psfig{file=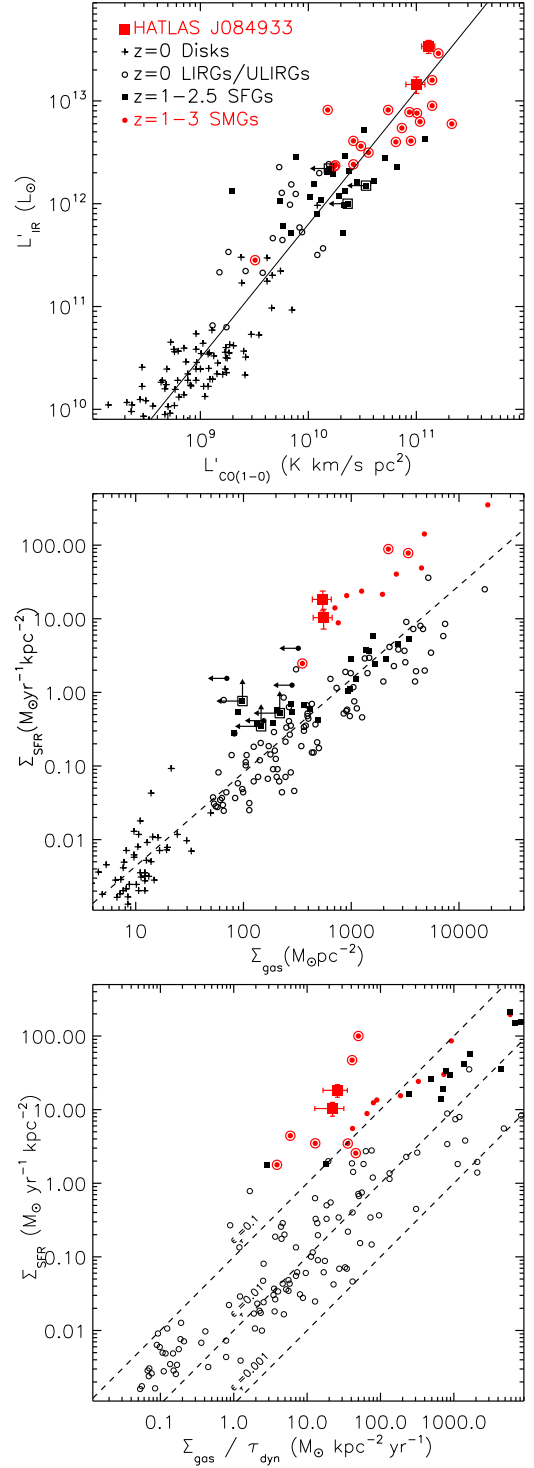,width=2.71in,angle=90}}
\caption{Plots relating the rate or surface density of star formation
  with the available gas. Components W and T are shown as red squares,
  with error bars, in each plot. Circled red symbols denote
  high-redshift measurements made directly in CO \jonezero. Low- and
  high-redshift star-forming galaxies (SFGs), e.g.\ $z=0$ disks, LIRGs
  and ULIRGs and $z=1$--2.5 BzK galaxies, are taken from
  \citet{genzel10} for all three plots.  {\it Top:} \lir\ versus
  $L^\prime_{\rm CO}$, where we find no compelling evidence that SMGs
  follow a different relation than the other star-forming
  galaxies. {\it Middle:} Surface-density S-K relation, where W and T
  lie amongst hyperluminous SMGs such as SMM\,J02399$-$0136 and
  GN\,20, offset from the relation for other star-forming
  galaxies. {\it Lower:} Elmegreen-Silk relation between
  star-formation surface density and the ratio of gas surface density
  and dynamical timescale.  W and T exhibit extreme star-formation
  efficiencies, even relative to SMGs. The fraction of their available
  gas converted into stars per dynamical timescale, $\epsilon_\tau$,
  is $>0.1$.}
\label{fig:sk}
\end{figure}

We begin this discussion in the upper panel of Fig.~\ref{fig:sk}: a
luminosity-luminosity plot of the observables, \lir\ and
$L^\prime_{\rm CO}$, for low- and high-redshift star-forming galaxies
(SFGs), e.g.\ $z=0$ disks, LIRGs and ULIRGs and $z=1$--2.5 BzK
galaxies, taken from \citet{genzel10}, and for SMGs. For this plot we
need make no assumption about the appropriate value of $\alpha_{\rm
  CO}$. The quality and depth of our continuum and CO data mean that
the errors bars for W and T can barely be discerned. Using
$L^\prime_{\rm CO}$ measurements made only in CO \jonezero, thereby
avoiding the need to adopt a \Tb\ ratio, alongside self-consistent
measurements of \lir, we find no compelling evidence that different
relations apply to SMGs and other SFGs, in line with the findings of
\citet{ivison11}. Here, we use the sample of SFGs from
\citet{genzel10} and find essentially the same result.

It is noteworthy that while component T contributes most to $I_{\rm
  CO}$ or $L^\prime_{\rm CO}$ in all three observed CO transitions,
component W is significantly brighter in the continuum at
IR-through-radio wavelengths. We find a similar variation between W
and T in relation to the far-IR/radio flux ratio, with \qir\ =
$2.30\pm 0.08$ and $2.53\pm 0.13$, respectively, spanning (and
consistent with) the tight correlation for star-forming galaxies
\citep[\qir\,$\approx 2.4$, e.g.][]{yun01}.  One might think their
different line-to-continuum ratios would lead to a relatively large
difference in their positions on the plot of \lir\ versus
$L^\prime_{\rm CO}$ (Fig.~\ref{fig:sk}, upper panel). In fact, W and T
both lie comfortably amongst the scatter of SMGs, suggesting that this
scatter may be intrinsic rather than due to measurement error.

Moving to the standard surface-density Schmidt-Kennicutt (S-K)
relation, $\sum_{\rm SFR}\propto \sum_{\rm gas}^N$, shown as the
middle panel of Fig.~\ref{fig:sk}, it is now necessary to adopt
value(s) for $\alpha_{\rm CO}$ appropriate for SFGs and
SMGs. \citet{genzel10} argue that 3.2\,\xunits\ is appropriate for
disks, regardless of redshift; our dynamical $\alpha_{\rm CO}$
constraints preclude the adoption of this value for W and T. We use
3.2 and 0.8\,\xunits\ for SFGs and SMGs, respectively, acknowledging
that this is overly simplistic\footnote{Changing $\alpha_{\rm CO}$ by
  a factor two (in opposite directions) for the SFGs and SMGs would
  bring them in line with one another. Because of the potential for a
  large systematic error in measurements of stellar mass, amongst
  other things, we view such a change as plausible.}
\citep[e.g.][]{narayanan12b}. We find W and T nestled amongst the
other SMGs again (Fig.~\ref{fig:sk}), but this time the SMGs are
offset\footnote{Any change in the area used to calculate the gas or
  SFR surface density results in a point shifting parallel to the
  power-law relation seen in this plot.}  significantly from the
relation for other SFGs.

The lower panel of Fig.~\ref{fig:sk} shows the Elmegreen-Silk (E-S)
relation between star-formation surface density and the ratio of gas
surface density and dynamical timescale\footnote{Although we have
  estimated the orbital timescale, in this calculation, we use the
  dynamical timescale of the disk, calculated using the half-light
  radius and the velocity dispersionm since \citet{krumholz12} show
  that use of the orbital time can artificially increase the apparent
  star-formation efficiency.}.  Taken at face value, W and T exhibit
extreme star-formation efficiencies. Several other SMGs are equally
extreme, converting $>10$\% of their available gas into stars per
dynamical timescale ($\epsilon_\tau> 0.1$). \citet{krumholz12} have
argued in favour of a simple volumetric star-formation law, linking
Galactic clouds, SMGs, and all scales between the two, with projection
effects leading to the observed scatter. Here we have discovered two
disks exhibiting star formation of such ferocity that they are capable
of converting $>10$\% of their molecular gas into stars with each
rotation. We require a number of parameters to conspire together if we
are to escape the conclusion that the gas in these disks does not obey
the simple volumetric star-formation law proposed by
\citet{krumholz12}.

Before leaving the S-K and E-S laws, we note that although \lir\ is an
effective tracer of $\sum_{\rm SFR}$ in Eddington-limited star-forming
disks, \citet*{ballantyne13} have argued that velocity-integrated CO
line intensity is a poor proxy for $\sum_{\rm gas}$. Resolved
observations of the mid-IR rotational lines of H$_2$ may provide the
ultimate probe of the star-formation relations in high-redshift disks,
requiring a far-IR space interferometer such as FIRI \citep{hi08}.

\subsection{Future evolution of the system}
\label{future}

What does the HATLAS\,J084933 system look like in the present day?

The extreme starbursts that brought this galaxy to our attention via
{\it Herschel} are likely to have been triggered by one or more
interactions \citep{engel10}. Certainly, C is distorted,
morphologically, and the distances between these galaxies are
consistent with periods of intense star formation during merger
simulations \citep[e.g.][]{springel05}.

However, without knowledge of the dark-matter halo they inhabit and
the transverse velocities of W, T, C and M, it is impossible to be
sure whether these galaxies are gravitationally bound, let alone
whether they will merge. We find that the total energy,
$1/2\,v^2+\Phi$\,(\kms)$^2$, of each of these systems is negative --
indicating that the galaxies are bound to one another -- if their
dynamical masses represent as little as 50\% of the total. This is
confirmed by $N$-body simulations, starting from the observed
velocities and positions on the sky, and assuming that the transverse
velocities do not exceed those in the line of sight. If these galaxies
inhabit an expected dark-matter halo of mass $\approx
10^{13}$\,M$_\odot$ \citep[e.g.][]{amblard11, hildebrandt13}, it is
difficult to escape the conclusion that the system is bound.

Taking another approach, we searched the $(500/h)^3$-Mpc$^3$ Millenium
Simulation at $z=2$ \citep{springel05}. In each friends-of-friends
halo we identified all of the sub-halos, as well as the most massive
sub-halo, which we take to be akin to W. We then computed the velocity
and spatial offset between the most massive sub-halo and all the other
sub-halos and tracked the $\approx 200$ cases where there are three
sub-halos within a radius of 300\,kpc (co-moving) with velocities
$>90$, $>600$ and $>800$\,\kms\ in projection (i.e.\ taking into
account likely line-of-sight distances and transverse
velocities). Those halos, typically $\approx 1.6 \times
10^{13}$\,M$_\odot$ at $z=2$, end up with a median mass of $\approx
3.8^{+5.5}_{-1.9}\times 10^{14}$\,M$_\odot$ in the present day (90\%
lower limit, $1.6\times 10^{14}$\,M$_\odot$) -- structures that are
thought to have a space density of a few $\times 10^{-7}$\,Mpc$^{-3}$
\citep[e.g.][]{reiprich02, vikhlinin09}.

\begin{figure}
\centerline{\psfig{file=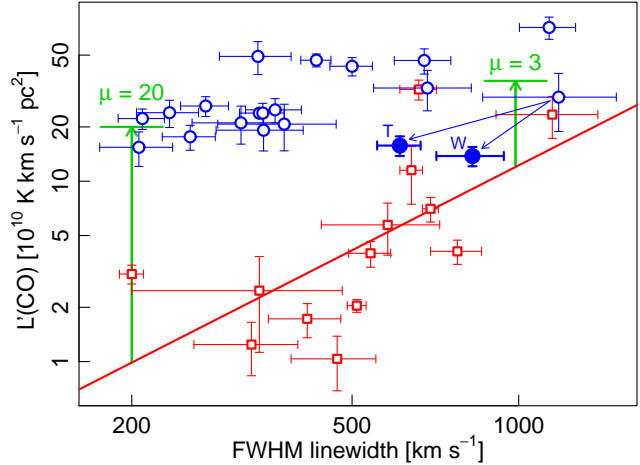,width=3.35in,angle=0}}
\caption{$L^\prime_{\rm CO}$ versus CO \jonezero\ FWHM line widths for
  SMGs with redshifts, fluxes and magnifications from the literature
  \citep[red squares -- for details, see][from whence this figure has
  been adapted]{harris12}. The power-law fit to these data represents
  an estimate of the intrinsic line luminosity versus line width
  relation for typical SMGs. Blue circles represent the lensed sample
  from \citet{harris12}. The point for HATLAS\,J084933 has been replaced
  with those appropriate for components W and T. W has moved well
  within the scatter of the intrinsic relation, with a predicted
  amplification consistent with unity. The predicted amplification for
  T is consistent with the lens model presented in \S\ref{lensmodel},
  $\mu = 1.5 \pm 0.2$.}
\label{fig:harris}
\end{figure}   

The scale over which we see the galaxies comprising HATLAS\,J084933,
$\approx$100\,kpc, is similar to the core radius of rich clusters of
galaxies, and the separation of W and T (in terms of kpc and \kms) is
typical of binary galaxies \citep[e.g.][]{turner76, schweizer87} and
within a factor $2\times$ of the projected separation of the two
dominant galaxies in the Coma cluster.\footnote{Coma (Abell 1656) is a
  B(inary)-type cluster, with two bright, supergiant galaxies, NGC
  4874 and NGC 4889, close to each other in the cluster core
  \citep{rood71}.} We conclude that such systems can survive for a
long time and are not necessarily in the process of merging. The four
starbursts in HATLAS\,J084933 may instead be signposting a
proto-cluster, where the majority of members are likely well below the
detection thresholds of our panchromatic imaging. It is ironic that
deliberate attempts to select clusters at submm wavelengths
\citep[targeting radio galaxies and quasars, for example
--][]{stevens03, debreuck04, greve07, priddey08} have met with only
limited success, while serendipitous discoveries are becoming
relatively commonplace \citep[e.g.][]{riechers10}.

\vspace*{0.3cm}
\subsection{On the selection of HyLIRGs, and their rarity}
\label{selection}

Having resolved HATLAS\,J084933 into its constituent parts, if we
place W and T individually on the \citeauthor{harris12} plot of
$L'_{\rm CO}$ versus FWHM line width (Fig.~\ref{fig:harris}) we find
that W has moved further within the scatter of the intrinsic relation,
with a predicted amplification close to unity. The predicted
amplification for T, $\mu\approx 2$, is consistent with the lens model
presented in \S\ref{lensmodel}.

We can estimate the space density of objects as rare as
HATLAS\,J084933 by utilising the semi-empirical model of
\citet{hopkins10} and \citet{hayward13}. We start with halo mass
functions and merger rates over the redshift range of interest derived
from the Millenium Simulation \citep{springel05}, assigning galaxies
to each halo following the abundance-matching techniques described in
\citet*{conroy09} to yield the galaxy-galaxy merger rate as a function
of mass, merger-mass ratio and redshift. Utilising the stellar mass
functions of \citet{marchesini09}, with extrapolations to higher
redshifts following \citet{fontana06}, we calculate the merger rate
for all galaxies with a stellar mass ratio equal to or greater than
that of W and T, with one component at least as massive as W.
Integrating the merger rates between $z=2$--6 ($t=2.4$\,Gyr), where we
are sensitive to such events via {\it H}-ATLAS, results in a space
density of approximately $10^{-7}$\,Mpc$^{-3}$, similar to the space
density predicted for extreme HyLIRGs \citep[][recall also
\S\ref{future}]{lapi11}. For the typical gas-consumption timescale
found in \S\ref{gasproperties}, $\tau_{\rm SMG} \approx$15--45\,Myr,
we then expect to find $\approx$3--9 such galaxies in the
$\approx$100\,deg$^2$ explored so far in {\it H}-ATLAS. The vast
majority of bright {\it Herschel} galaxies have been shown to be
lensed \citep{negrello10}, but the low end of this range is
plausible. This suggests that the most luminous phase of these
galaxies has a short duration, $\ls 15$\,Myr, and/or that such systems
require some other rare property -- counter-rotating disks being one
possibility. Stellar mass functions at high redshift have significant
uncertainties associated with them; nonetheless, this calculation
emphasises the rarity of major mergers such as that between W and T
and illustrates the efficiency of the method employed here to find
extreme HyLIRGs. The observed surface density of these systems is such
that even with short durations for their most luminous phases, they
can end up as a subset of the massive current-day clusters discussed
in \S\ref{future}.

HATLAS\,J084933 was selected on the basis of its broad CO line profile
as a low-magnification, intrinsically luminous system, so it is ironic
that its wide line profile is in part due to the multiple nature of
the source and that future searches for single maximal-luminosity
galaxies may require the selection of less extreme profiles.  However,
fundamentally we conclude that the method explored here for the
selection of the most extreme starburst events in the Universe shows
promise.

\section{Conclusions}
\label{conclusions}
 
\begin{enumerate}

\item Exploiting the relationship between CO luminosity and line width
  determined for unlensed starbursts, we have identified and removed
  gravitationally lensed sources from the brightest galaxies found in
  the widest extragalactic {\it Herschel} survey to yield a sample of
  intrinsically luminous galaxies. Here, we report deep panchromatic
  follow-up observations of the best candidate HyLIRG
  system, HATLAS\,J084933, with sub-arcsecond spectral and spatial
  resolution, which have led to the discovery of at
  least four starbursting galaxies across a $\approx$100-kpc region at
  $z=2.41$, each with a significant mass of stars.

\item The two brightest galaxies, W and T, are separated by
  $\sim$85\,kpc on the sky. W suffers no gravitational amplification.
  T is marginally lensed by a foreground lenticular. 

\item The panchromatic SEDs of W and T reveal that both are associated
  with high-luminosity events -- HyLIRGs. If contributions to \lir\
  from AGN are small, and to date we have found no reason to suspect
  otherwise, then W and T have $\log L_{\rm IR} \sim 13.52$ and 13.16,
  \Td\ = 36--40\,\kelvin, and are generating $\approx$3400 and
  $\approx$1500\,M$_{\odot}$\,yr$^{-1}$ of stars, respectively.

\item Two other gas-rich galaxies, M and C, detected nearby, make
  HATLAS\,J084933 reminiscent of the first SMG found and the
  best-studied HyLIRG to date, SMM\,J02399$-$0136, which comprises a
  complex galactic nursery with three massive stellar components, each
  seen during a different evolutionary stage (Ferkinhoff et al., in
  preparation).  The four galaxies known to comprise HATLAS\,J084933
  are most likely bound. $N$-body simulations and comparison with the
  Millenium Simulation suggest that in the present day the system may
  resemble a B(inary)-type cluster of mass, $\approx
  10^{14.6}$\,M$_\odot$.

\item Sub-arcsecond interferometric imaging reveals that the two
  brightest galaxies span $\sim$3\,kpc FWHM in CO \jfourthree, and in
  submm and radio continuum, and roughly double that in CO \jonezero.
  Alongside the detection of \Cplus, this supports a scenario
  involving a widespread burst of intense star formation.

\item Exquisite 3-D imaging from JVLA and IRAM PdBI reveal
  counter-rotating gas disks in W and T, a scenario that has long been
  predicted to lead to the most intense bursts of star formation. This
  suggests that similarly luminous galaxies will often be found in
  pairs. Their modest velocity dispersions mean that the disks are
  prone to instabilities on many scales, even that of the disks
  themselves; they are undergoing coherent, extreme starbursts, at
  close to the Eddington limit, which are not confined to their
  nuclei. Despite their disk-like morphologies and dynamics, their
  specific star-formation rates place them $\sim5\times$ above the
  main sequence.

\item W and T have CO line intensities and widths  typical of the
  brightest SMGs, with slightly higher \Tb\ ratios. Three independent
  estimates of the CO-to-H$_2$ conversion factor -- exploiting the
  dynamical mass, the mass of dust and a recent metallicity-dependent
  relation -- suggest $\alpha_{\rm CO} = 0.8\pm 0.4$\,\xunits\ and
  hence gas masses in W and T of $\approx 10^{11}$\,M$_\odot$. Their
  gas fractions, $\sim40$\%, are difficult to reconcile with
  cosmological simulations.

\item We have determined that the single-dish CO line profile of
  HATLAS\,J084933 is exceptionally broad because W and T have
  relatively wide lines, broadened further in the \citet{harris12}
  integrated GBT spectrum by the modest velocity offset between them,
  and by contributions from a number of faint CO clumps slightly
  redward of the two dominant components. We can thus
  reconcile the GBT CO line profile with that of the various galaxies
  found to comprise HATLAS\,J084933. These situations likely happen
  only rarely, even in the widest extragalactic {\it Herschel}
  surveys. We estimate that only a few systems of this nature will be
  found in every 100\,deg$^2$ surveyed.

\item Placing W and T individually on the \citeauthor{harris12} plot
  of $L'_{\rm CO}$ versus FWHM CO line width, W and T move further
  within the scatter of the intrinsic relation, with a predicted
  amplification close to unity. We conclude that the method
  outlined here for the selection of extreme starbursts can indeed
  separate HyLIRGs from the more numerous, less luminous, strongly
  lensed population of IR-bright galaxies, and may also pinpoint some
  distant clusters of starbursting proto-ellipticals.

\item Plotting \lir\ versus $L^\prime_{\rm CO}$, W and T lie amongst
  the SMGs on a single relation that is consistent with all
  star-forming galaxies. They also lie amongst the SMGs on the classic
  Schmidt-Kennicutt and Elmegreen-Silk plots, but the SMGs now follow
  a relation that is distinct from those of other star-forming
  galaxies. Their star-formation efficiency is significantly higher,
  and is difficult to reconcile with a simple volumetric
  star-formation law in which $\sim$1\% of the available gas is
  converted into stars during each local free-fall time.
\end{enumerate}

Understanding systems as luminous as HATLAS\,J084933 represents a
challenge. However, the observations presented here have shed
considerable light on the characteristics that likely lead to the
formation -- however briefly -- of the most luminous star-forming
galaxies in the Universe. Observations scheduled for Cycle 1 with the
Atacama Large Millimeter/Submillimeter Array promise to reveal more
about this dramatic and complex proto-galactic environment, probing
the structure of the disks and sensitive to the presence and influence
of buried AGN.

\section*{Acknowledgements}

RJI acknowledges support from the European Research Council (ERC) in
the form of Advanced Grant, {\sc cosmicism}. The {\it Herschel}-ATLAS
is a project with {\it Herschel}, which is an ESA space observatory
with science instruments provided by European-led Principal
Investigator consortia and with important participation from NASA. The
{\it H}-ATLAS website is {\tt www.h-atlas.org}. US participants in
H-ATLAS acknowledge support from NASA through a contract from JPL. IRS
acknowledges support from STFC and ERC. MN and GdZ acknowledge
financial support from ASI/INAF agreement I/072/09/0.

{\it Facilities:} \facility{CARMA, CSO, Herschel, Hubble Space
  Telescope, IRAM PdBI, JVLA, Keck, SMA, Spitzer, VISTA.}

\bibliographystyle{apj}
\bibliography{rji}

\end{document}